\documentclass[ reprint,
 amsmath,amssymb,amsart,aps,pra,noshowpacs
]{revtex4-1}

\newcommand*{\new}{\textcolor{black}} 
\usepackage{graphicx}
\usepackage{dcolumn}
\usepackage{bm}
\usepackage[colorlinks=true,urlcolor=blue,citecolor=blue,linkcolor=blue]{hyperref}

\usepackage{mathtools}
\usepackage{epstopdf}

\usepackage[dvipsnames]{xcolor} 
\usepackage[english]{babel}
\usepackage[utf8]{inputenc} 
\usepackage{amsmath} 
\usepackage{braket} 
\usepackage[load=abbr]{siunitx} 
\usepackage{tikz}
\usepackage{tabularx}
\usepackage{todonotes}
\usepackage{tikz} 
\usetikzlibrary{calc}

\begin{document}

\preprint{APS/123-QED}

\title{Strong zero-field F\"orster resonances in K-Rb Rydberg systems}
\author{J.~Susanne~Otto}\email{susanne.otto@postgrad.otago.ac.nz}
\author{Niels~Kj{\ae}rgaard}
\author{Amita~B.~Deb}

%
 \email{amita.deb@otago.ac.nz}
\affiliation{QSO---Centre for Quantum Science , Dodd-Walls Centre for Photonic and Quantum Technologies, and the Department of Physics, University of Otago, Dunedin, New Zealand.}

\begin{abstract}
We study resonant dipole-dipole coupling and the associated van der Waals energy shifts in Rydberg excited atomic rubidium and potassium and investigate F\"orster resonances between interspecies pair states. A comprehensive survey over experimentally accessible pair state combinations reveals multiple candidates with small F\"orster defects. We crucially identify the existence of an ultrastrong, ``low" electric field K-Rb F\"orster resonance with a extremely large zero-field crossover distance \new{exceeding \SI{100}{\mu m} between the van der Waals regime and the resonant regime. This resonance allows for a strong interaction over a wide range of distances and by investigating its dependence on the strength and orientation of external fields we show this to be largely isotropic. As a result, the resonance offers a highly favorable setting for studying long-range resonant excitation transfer and entanglement generation between atomic ensembles in a flexible geometry. The two-species K-Rb system establishes a unique way of realizing a Rydberg single-photon optical transistor with a high-input photon rate and we specifically investigate an experimental scheme with two separate ensembles. }


\end{abstract}
\maketitle
\section{INTRODUCTION}
\vspace{-3mm}
In recent years, ultracold Rydberg atoms have emerged as a prominent resource for numerous quantum-enabled technologies including quantum information processing \cite{Saffman2010}, quantum simulation \cite{Weimer2010, Altman2019}, quantum nonlinear optics \cite{Firstenberg2016}, and hybrid quantum devices \cite{Lauk2020}. Arrays of Rydberg atoms, for example, have been used for realizing high-fidelity quantum gates \cite{Isenhower2010, Saffman2016} and for generating strongly correlated phases of many-body quantum systems \cite{Schaus2012, Keesling2019, Omran2019,Bernien2017,Labuhn2016}, and strong photon-photon interactions mediated by Rydberg atoms \cite{Pritchard2010,Peyronel2012} have been exploited to realize single-photon switches and transistors \cite{Tiarks2014,Gorniaczyk2016}, quantum memories \cite{Li2016a} and photonic phase gates \cite{Tiarks2016}. The signature feature of ultracold Rydberg atoms is their strong dipole-dipole interactions which gives rise to two important mechanisms. The first is the blockade effect where a single Rydberg excitation from laser light forbids further excitations within a certain distance from the first due to the energy shift caused by the Rydberg-Rydberg interaction. In an atomic ensemble, this leads to a single collective atomic excitation shared among the atoms resulting in an effective two-level ``superatom" \cite{Weber2015}. The second important mechanism is that the long-range interaction, when resonant, can cause coherent excitation transfer between atoms that are far apart \cite{Ditzhuijzen2008, Ravets2014, Anderson1998,Gunter2013}. The strength and range of the dipole-dipole interaction are the most critical factors in the practical implementation of these two effects that underpin a vast range of experimental observations with ultracold Rydberg systems to date.
\begin{figure}[t]
	\includegraphics[clip, trim= 0.0cm 0.0cm 0cm 1.1cm,width=0.99\columnwidth]{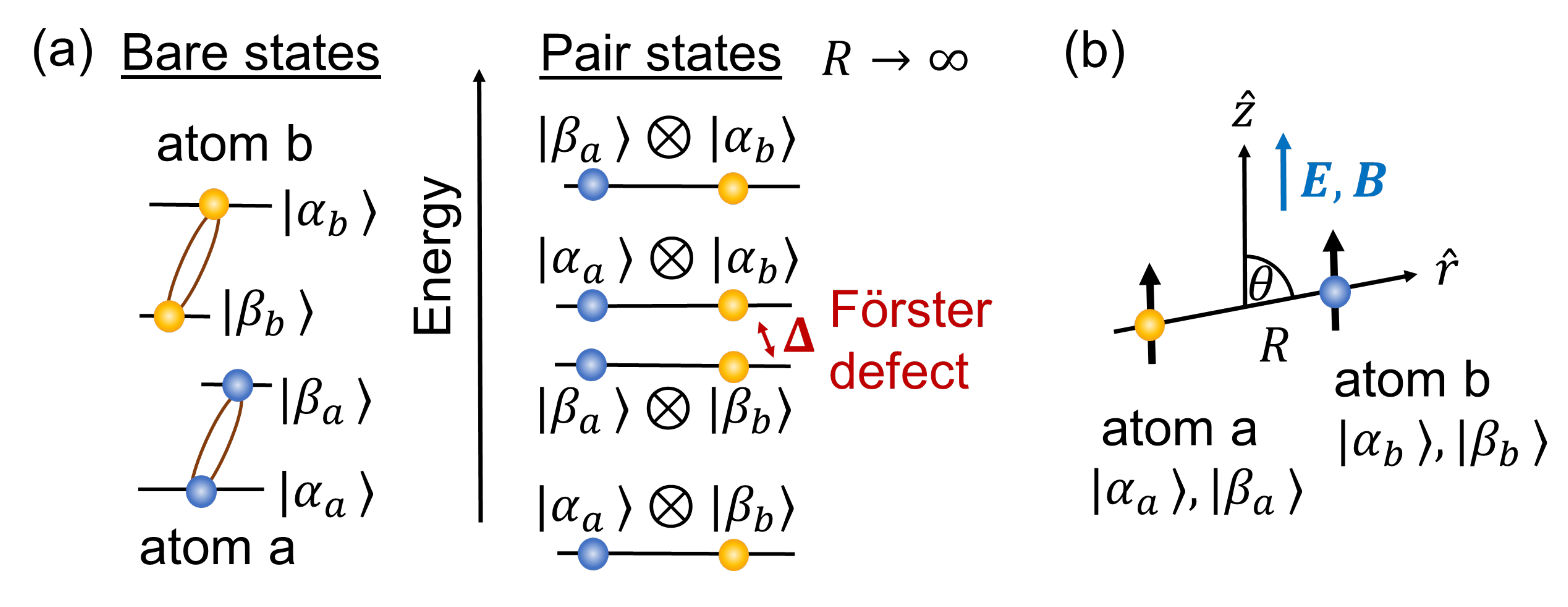}%
	\caption{(a) Pair state energies for two atoms a and b in Rydberg bare states $\ket{\alpha_a}$, $\ket{\beta_a}$, and $\ket{\alpha_b}$, $\ket{\beta_b}$ with F\"orster defect $\Delta = E(\alpha) - E(\beta)$. (b) Two Rydberg atoms with dipole moments $\bf{a}$ and $\bf{b}$ at an interatomic separation $R$. The angle $\theta$ is determined by the quantization axis $\hat{z}$ and the interatomic axis $\hat{r}$. $\ket{\alpha_{a}}$, $\ket{\alpha_{b}}$, $\ket{\beta_a}$ and $\ket{\beta_{b}}$ represent sets of quantum numbers $\ket{n l j m_j}$, consisting of the principal quantum number $n$, the orbital angular momentum $l$, the total electronic angular momentum $j$ and the $\hat{z}$ projection of the angular momentum $m$. For the case of external fields, $\bf{B}$ and $\bf{E}$, the field direction defines the quantization axis, otherwise we define $\hat{z}$ directed along $\hat{r}$.}
	\label{fig0}
\end{figure}
\newline
Denoting by $\ket{\alpha_i}$ and $\ket{\beta_i}$ the initial and final states of a Rydberg atom $i$, the coherent dipolar coupling between two pairs of Rydberg states $|\alpha \rangle= |\alpha_a,\alpha_b\rangle$ and $|\beta\rangle = |\beta_a,\beta_b\rangle$ in a pair of atoms $a$ and $b$ is of the form ${\cal{U}} \sim C_3/R^3$, where $R$ is the distance between the two atoms (Fig.~\ref{fig0}) and $C_3$ is a constant that depends on the dipole matrix elements and the orientation of the dipoles with respect to the interatomic axis \cite{Saffman2010, Weber2017}. The energy eigenstates of the system are determined by the relative magnitudes of ${\cal{U}}$ and the energy difference between the two pair states $\Delta = E(\alpha) - E(\beta)$ at infinite separation, also known as the F\"orster defect \cite{Zeros_Rydberg}. In the regime where $\cal{U} \ll$ $ |\Delta|$, the dominant effect from the state $|\beta \rangle $ on an initially excited state $|\alpha \rangle$ is an energy shift that is second order in $\cal{U}$ and has the van der Waals form $ {-\cal{U}}^2/\Delta \sim -C_6/R^6$. In the regime where ${\cal{U}} \gtrsim |\Delta|$, the dipolar coupling is resonant and the energy shift varies much more slowly with distance as $\sim 1/R^3$. A F\"orster resonance occurs when the F\"orster defect between the pair states vanishes, such that the coupling is resonant and the energy shift has a $1/R^3$ scaling for arbitrary distances. In practice, an external dc or microwave electric field is used to tune a naturally occurring small F\"orster defect to zero \cite{Afrousheh2004, Bohlouli-Zanjani2007}. 
F\"orster resonances are a highly useful tool in ultracold Rydberg physics, as they enable fast external control of the strength and angular variation of Rydberg interactions \cite{Ravets_2015}, extend the range of Rydberg blockade by allowing a relatively slow $1/R^3$ fall-off, and can realize long-distance dipolar exchange of states between atoms or atomic ensembles \cite{Paris-Mandoki2016}. Indeed, F\"orster resonances have been widely studied in the context of dipole blockade \cite{Vogt2006, Ryabtsev2010}, excitation hopping between single atoms and atomic ensembles \cite{Ravets2014, Ditzhuijzen2008}, non-destructive imaging of Rydberg atoms \cite{Gunter2013}, enhanced gain of single-photon optical transistors \cite{Tiarks2014,Gorniaczyk2016}, and non-demolition quantum-state measurements of Rydberg atom qubits \cite{Beterov2015}. 

Experimental and theoretical studies on quantum processing and quantum optical tools based on Rydberg systems have so far almost exclusively focused on using a single atomic species. Advanced atom trapping in geometries such as one-, two-, and three-dimensional arrays of traps for single atoms \cite{Bernien2017, Barredo2018} and atomic ensembles \cite{Weber2015, Chisholm2018, Roberts2014}, however, enable configurations of spatially separated clouds of different species. 
A two species system of rubidium and cesium (Rb-Cs) atoms has been recently proposed as a promising approach for qubit systems with suppressed cross-talk between computational and measurement qubits \cite{Beterov2015, Zeng2017}. Dimer states and dispersion coefficients in K-Rb Rydberg systems have also been theoretically studied \cite{Samboy2017,Olaya2019}. K-Rb in particular offers the possibility to explore effects due to the bosonic ($^{85}$Rb, $^{87}$Rb, $^{39}$K, $^{41}$K) and the fermionic ($^{40}$K) quantum statistics of the atoms.
In this paper, we study F\"orster resonances arising in the dipole-dipole interaction between rubidium and potassium atoms in their Rydberg states. We observe a number of near F\"orster resonances with fortuitously small zero-field F\"orster defects resulting in a $1/R^3$ scaling in energy shift for distances up to 100\,$\mu$m. Crucially, these near resonances can further be brought to exact resonance by applying very small electric fields ($< \SI{50}{\mV/cm}$). This is significant because high-lying Rydberg states possess giant electric polarizabilities and consequently the application of even moderate electric fields can mix many Rydberg states at small distances, compromising their suitability for Rydberg blockade. \new{Our investigation covers F\"orster resonances for pair states comprised of $s$ and $d$ angular momentum states that can be excited from atomic ground states via two-photon excitations. We provide pair interaction potentials and angular dependencies for the most relevant pair state, which has a strong and largely isotropic interaction over a wide range of distances. Importantly, we describe its potential use in a single-photon transistor that capitalizes on the use of spatially separated rubidium and potassium ensembles.} \newline
The paper is organized as follows. In Sec.~\ref{sec2}, we provide a brief overview of dipole-dipole interactions between Rydberg states and provide general formulas for the calculation of characteristic interaction parameters such as interaction coefficients and crossover distances. In Sec.~\ref{sec3}, we analyze in detail the F\"orster resonances for a range of K-Rb pair states on a grid of principal quantum numbers $50 \leq n_\text{K},n_\text{Rb} \leq 110$ for a variety of angular momentum states and tabulate the most promising F\"orster resonances in this system. In addition, we compare them to the known resonances in Rb-Rb, Rb-Cs and Cs-Cs systems. For the state with the smallest F\"orster defect, we investigate in Sec.~\ref{sec4} the interaction potentials in the presence of external fields, and discuss the angular dependence of the interaction. In Sec.~\ref{sec5}, we explore benefits of the strong F\"orster resonances for possible applications in photonic devices and analyze a Rydberg single-photon optical transistor. We summarize our results in Sec.~\ref{sec6}.
\noindent
\section{DIPOLE-DIPOLE INTERACTIONS}
\label{sec2}
We consider two Rydberg atoms $a$ and $b$ with an interatomic separation $R \gg r_\text{LR}$ where $r_\text{LR} = \sqrt{\langle s_1 \rangle^2 + \langle s_2 \rangle^2}$, $\langle s_i \rangle$ characterizes the spatial extent of the electronic cloud for an atom in the Rydberg state. In this regime, the electron clouds are non-overlapping and the two atoms interact via electrostatic interaction that can conveniently be expressed using a multipole expansion. The dominant term in the expansion is typically the dipole-dipole interaction \cite{Zeros_Rydberg}, 
\begin{equation}\label{vddx}
V_\text{dd}(\mathbf{R}) = \frac{1}{4\pi\epsilon_0} \frac{ \mathbf{a} \cdot \mathbf{b} - 3 (\mathbf{a} \cdot \mathbf{\hat{r}})(\mathbf{b}\cdot \mathbf{\hat{r}}) }{R^3},
\end{equation}
where $\mathbf{\hat{r}}$ is a unit vector along the interatomic axis and $\mathbf{a} = e\, \mathbf{r_a}$ and $\mathbf{b} = e\, \mathbf{r_b}$ are the dipole moments of the two atoms. The overall interaction is described by the total Hamiltonian 
\begin{equation}
H = H_a \otimes \mathrm{1}_b + H_b \otimes \mathrm{1}_a + V_\text{dd},
\label{ham}
\end{equation}
consisting of the single-atom Hamiltonians $H_a$ and $H_b$, identity operators $\mathrm{1}_a$ and  $\mathrm{1}_b$, and the interaction operator $V_\text{dd}$, the quantum mechanical equivalent to the classical case in Eq.~(\ref{vddx}). The eigenstates and eigenvalues of the total Hamiltonian at atomic separations $R$ can be used to compose potential landscapes of the interaction. The initial states $\ket{\alpha_a}$ and $\ket{\alpha_b}$ of atom $a$ and $b$ form a pair state $\ket{\alpha_a \alpha_b}$ which couples to pair combination $\ket{\beta_a \beta_b}$ via $\braket{\alpha_a \alpha_b|V_\text{dd}|\beta_a \beta_b}$, see Fig.~\ref{fig0}(a), and leads to off-diagonal terms in the full Hamiltonian $H$. The main task is therefore to calculate these off-diagonal matrix elements which is best done by describing the dipole moments in the spherical basis defined by $a_0 = a_z, a_{\pm 1} = \mp \frac{a_x \pm i a_y}{\sqrt{2}}$ \cite{Sakurai2014} for atom $a$ and similarly for atom $b$. We transform Eq.~(\ref{vddx}) to \cite{Paris-Mandoki2016}
\begin{align}
 \nonumber V_\text{dd} = \frac{1}{4\pi \epsilon_0 R^3} \Bigl[ \frac{(1-3\text{cos}^2\theta)}{2}(2 a_0 b_0 + a_{+1}b_{-1}+a_{-1}b_{+1}) \\
\nonumber -\frac{3\text{sin}\theta\text{cos}\theta}{\sqrt{2}} (a_{-1}b_0-a_{+1}b_0+a_{0}b_{-1}-a_{0}b_{+1})\\
- \frac{3\text{sin}^2\theta}{2}(a_{+1}b_{+1}+a_{-1}b_{-1}) \Bigr].
\label{eqrad}
\end{align}
In the absence of external fields, we choose the quantization axis, $\mathbf{\hat{z}}$, to line up with the interatomic axis, $\mathbf{\hat{r}}$. Otherwise, the external fields define the direction of the quantization axis and thereby the angle $\theta$, as depicted in Fig.~\ref{fig0}(b).
\begin{figure*}
\includegraphics[clip, trim= 3.5cm 0.0cm 2.5cm 0.5cm, width=0.99\textwidth]{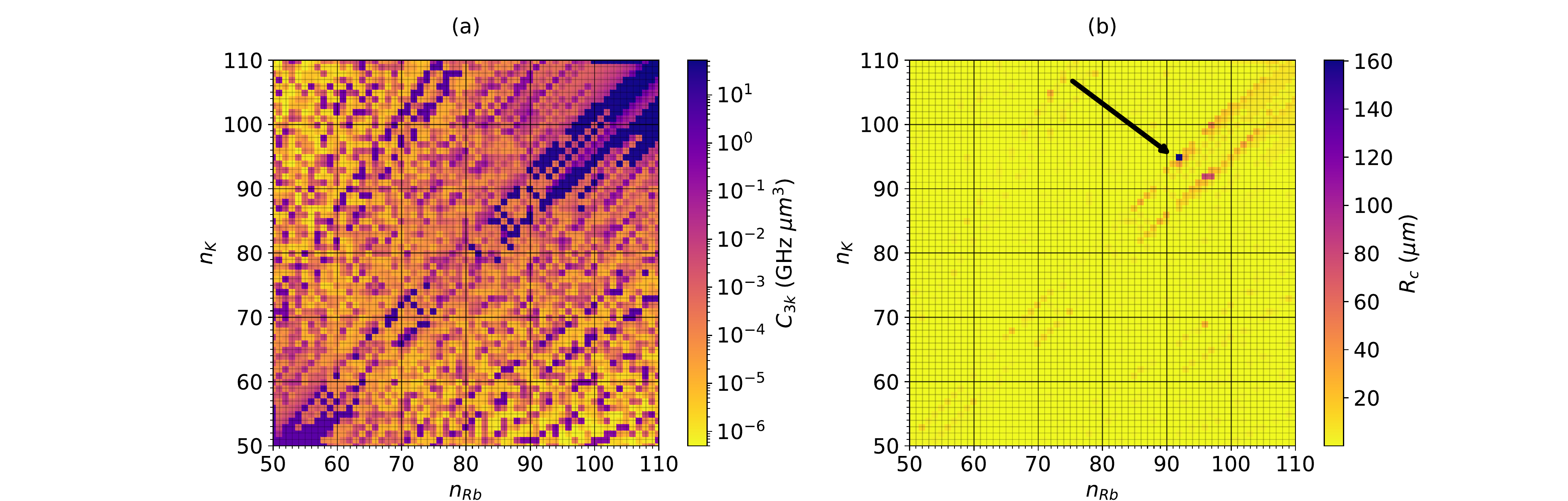}%
\caption{Interaction coefficients $|C_{3k}|$ (a) and corresponding crossover distances $R_c$ (b) of K-Rb pair states for their lowest F\"orster defects $|\Delta|$ and for a range of combinations of principal quantum numbers 50 $\leq n_\text{Rb}, n_\text{K} \leq$ 110 for initial $\ket{n_\text{Rb}s_{1/2}\;n_\text{K}s_{1/2}}$ states. A large crossover distance $R_c = (C_{3k}^2\mathcal{S}_k/(\hbar^2\Delta^2))^{1/6}$ characterizes states with small $\Delta$ and large $C_{3k}$. The figure shows the maximum crossover distances that can be obtained for a channel $k$ by choosing the angular momentum channel with largest $\mathcal{S}_k$.}
\label{figS_state} 
\end{figure*}
Numerous Rydberg pair states can be coupled to an initial pair state via the dipole-dipole interaction, but in practice, only a few of the pair states are sufficiently close to each other (small $\Delta$) to interact strongly. From Eq.~(\ref{eqrad}) it becomes clear that one has to calculate the dipole matrix elements $\braket{\alpha_a \alpha_b|a_q b_{q'}|\beta_a \beta_b}$, where $ q,q' \in {0,\pm 1}$. These all involve products of $\langle \alpha_a | a_q |\beta_a \rangle$ and $\langle \alpha_b | b_{q'} |\beta_b \rangle$, which are themselves matrix elements of spherical tensor operators with respect to angular-momentum states such as $\ket{\alpha_a} = \ket{n_\alpha,l_\alpha,j_\alpha,m_\alpha}$. The Wigner-Eckart theorem \cite{Sakurai2014} allows one to simplify the tensor matrix elements, resulting in 
\begin{equation}
\langle \alpha_a | a_q |\beta_a \rangle = C^{j_{\alpha_a},m_{\alpha_a}}_{j_{\beta_a} 1,m_{\beta_a}q}\frac{\langle j_{\alpha_a} ||e r_a ||j_{\beta_a} \rangle}{\sqrt{2j_{\beta_a} +1}}.
\label{WE}
\end{equation}
Here, $\langle j_{\alpha_a} ||e r_a||j_{\beta_a} \rangle$ is a reduced matrix element in the fine structure basis of atom $a$, $r_a$ is the internal position operator of atom $a$ and $C^{j_{\alpha_{a}},m_{\alpha_a}}_{j_{\beta_a} 1,m_{\beta_a}q}$ are Clebsch-Gordan (CG) coefficients. The reduced matrix element can be expressed by the radial wave functions $\mathcal{R}_{\alpha_a}(r)$ and $\mathcal{R}_{\beta_a}(r)$ of atom $a$ and $b$, 
\begin{align}
\nonumber  \langle j_{\alpha_a} ||e r_a ||j_{\beta_a} \rangle = (-1)^{(2l_{\beta_a} + s+ j'_{\alpha_a}+1)}  \\
\nonumber \times \sqrt{(2 j_{\alpha_a})(2j_{\beta_a}) (2 l_{\alpha_a} + 1)(2l_{\beta_a})} \\
\nonumber \times \int_{0}^{\infty} \mathcal{R}_{\alpha_a}(r) e r \mathcal{R}_{\beta_a}(r) r^2 \text{dr} \\
\nonumber \times \underbrace{\begin{Bmatrix}
j_{\alpha_a} & 1 & j_{\beta_a} \\
l_{\beta_a} & s & l_{\alpha_a} \\
\end{Bmatrix}}_\text{Wigner 6-\textit{j} symbol}
\underbrace{\begin{pmatrix}
l_{\alpha_a} & 1 & l_{\beta_a} \\
0 & 0 & 0 \\
\end{pmatrix}}_\text{Wigner 3-\textit{j} symbol},
\end{align}
and most importantly, it is completely independent of the magnetic quantum numbers $m_{\alpha_a}$, $m_{\beta_a}$, and the light polarization $q$ which are determined by the orientation of the two atoms with respect to the quantization axis. The commonly used CG coefficients describe the angular coupling of $\ket{\alpha_a}$ and $\ket{\beta_a}$. Only the nonzero matrix elements have to be considered, which are those satisfying conservation of angular momentum, $m_{\beta_a} = m_{\alpha_a} + q$.
Hence, when calculating the matrix elements for the dipole-dipole interaction we only need to calculate the reduced matrix element once, since it is identical for all terms in Eq.~(\ref{eqrad}). 
\newline
\newline
If we only consider a single channel $k$ which connects two pair states, we can express the matrix elements of the dipole-dipole interaction as
\begin{eqnarray}
\langle \alpha | V_\text{dd} |\beta \rangle = \frac{C_{3k}}{R^3} {\mathcal{S}_{k}}^{1/2},
\label{ceq2}
\end{eqnarray} 
where we define the interaction coefficient 
\begin{eqnarray}
C_{3k} = \frac{e^2}{4\pi\epsilon_0}\frac{\langle j_{\alpha_a} ||r_a ||j_{\beta_a} \rangle \langle j_{\alpha_b} ||r_b ||j_{\beta_b} \rangle }{\sqrt{(2j_{\beta_a} + 1)(2j_{\beta_b} + 1) }}
\label{ceq}
\end{eqnarray} 
which contains the reduced matrix element for both atoms. This spin-independent coefficient determines the interaction of two pair states with quantum numbers $(n_{\alpha_a},n_{\alpha_b},l_{\alpha_a},l_{\alpha_b},j_{\alpha_a},j_{\alpha_b}) \leftrightarrow (n_{\beta_a},n_{\beta_b},l_{\beta_a},l_{\beta_b},j_{\beta_a},j_{\beta_b})$. \new{The angular dependence of the interaction is summarized in the coefficient $\mathcal{S}_k^{1/2}$ \cite{Beterov2015}, which consists of CGs that additionally depend on the quantum numbers $m_{\alpha_a}$, $m_{\alpha_b}$, $m_{\beta_a}$, $m_{\beta_b}$, and $q$.}
\newline
\newline
The total Hamiltonian in the pair-state basis reduces to a standard two level Hamiltonian
\begin{equation}
\label{twolevel1}
\begin{pmatrix}
0 & \frac{C_{3k}}{R^3} {\cal{S}}_k^{1/2} \\
\frac{C_{3k}}{R^3} {\cal{S}}_k^{1/2} & \hbar \Delta \\
\end{pmatrix},
\end{equation}
yielding an interaction energy \cite{Beterov2015}
\begin{equation}
\mathcal{U}_k(R) = \frac{\Delta}{2} \left( 1 - \sqrt{1 + \frac{4 C_{3k}^2 \mathcal{S}_k}{\hbar^2 \Delta^2 R^6}}\, \right), 
\label{Ufull}
\end{equation}
for the state that adiabatically connects to the initial pair state.
\noindent
The crossover distance
\begin{equation}
R_c = \left(\frac{C_{3k}^2 \mathcal{S}_k}{\hbar^2\Delta^2}\right)^{1/6},
\label{Cdistance}
\end{equation} 
where  $\V_\text{vdW}(R_c) =  \hbar \Delta$, defines the boundary between the resonant $1/R^3$ dipole-dipole regime and the $1/R^6$ vdW regime with $V_\text{vdW}(R)=-\frac{C_{3k}^2 \mathcal{S}_k}{\hbar \Delta} \frac{1}{R^6}$. The sign of the F\"orster defect $\Delta$ determines if the interaction is attractive ($\Delta > 0)$ or repulsive $(\Delta < 0)$. While the angular momentum factors $\mathcal{S}_k$ are species independent, we have to take into account that for interspecies interaction the atoms are distinguishable at all times even if atoms a and b are initially sharing the same state quantum numbers. We note that for a single-species system with an initial pair state $\ket{\alpha_a\alpha_a}$ the angular momentum factor is twice as large due to the equal coupling to $\ket{\beta_a\beta_b}$ and $\ket{\beta_b\beta_a}$.  

\section{K-Rb F\"ORSTER RESONANCES}
\label{sec3}
To identify K-Rb pair states with particularly long-range interaction we assume that a single angular momentum channel $k$ is dominant, i.e. only one final pair state has to be considered. This presupposes that all other pair states are weakly coupled or have large F\"orster defects. Following the convention in  \cite{Beterov2015}, we calculate F\"orster defects $\Delta$ and coefficients $C_{3k}$ expressed in the fine structure basis, Eq.~(\ref{ceq}). There are generally many possible combinations of Rydberg levels that can be employed for a F\"orster process due to the high density of Rydberg states for high-$n$ (energy spacing between adjacent $n$ values scales as $n^{-3}$). We shall focus our attention on initial $s$ and $d$ pair states, since these are the most convenient to produce in optical schemes using laser light. Starting with atoms in their ground state $s_{1/2}$, Laporte's electric dipole selection rule, $\Delta l = \pm1$, only allows for one-photon excitations to excited $p$ states. For excitation to a Rydberg state, such one-photon transitions demand laser light in the UV spectrum for $40 \lesssim n \lesssim 150$, which is not commonly available in laboratories. Instead, a two-photon transition can be used for the excitation into a high $ns$ or $nd$ Rydberg state, via an intermediate $p$ state. Common two-photon excitation schemes involve combinations of infrared and blue radiation \cite{RbEIT,Arias:17}, allowing for implementations with readily available laser systems. Adding a second laser field and an intermediate third atomic level additionally enables the application of coherent phenomena as electromagnetically induced transparency (EIT) \cite{RevModPhys.77.633}. 
\newline
For an initial Rydberg pair state $\ket{n_{\rm Rb}\ell^{\rm Rb}_{j^{\rm Rb}}}\otimes \ket{n_{\rm K}\ell^{\rm K}_{j^{\rm K}}}$, we identify the complementary pair state $\ket{n'_{\rm Rb}\ell^{'\rm Rb}_{j^{'\rm Rb}}}\otimes \ket{n'_{\rm K}\ell^{'\rm K}_{j^{'\rm K}}}$ that achieves the smallest F\"{o}rster defect and obeys Laporte's rule. Once the complementary pair state that minimizes the F\"{o}rster defect is found, we calculate its interaction coefficient $C_{3k}$ with respect to the initial state pair. We explore the variation of such defect-minimized $C_{3k}$ coefficients on a grid of initial principal quantum $50\le n_{\rm Rb},n_{\rm K}\le110$ and present our results for initial $l=s$ and $l=d$ states in the following sections. 
\subsection{$s_{1/2} + s_{1/2}$ channels}
\label{sec2a}
\begin{table*}[t]
\caption{\label{table1}%
Pair states with small F\"orster defects $\Delta$ and high $C_{3k}$ coefficients, corresponding to Fig.~\ref{figS_state} and \ref{fig_d}, allowing for large $R_c$. \new{For initial $s+s, d+d, s+d$ and $d+s$ pair states we list the combinations with $R_c > \SI{55}{\mu m}$.} More pair combinations can be found in Table~\ref{table_s} in Appendix~\ref{C} . 
}
\begin{ruledtabular}
\renewcommand{\arraystretch}{1.4}
\begin{tabular}{p{1.3cm}p{1.1cm}ccc}
\textrm{Initial}&
\textrm{Final}&
\textrm{$C_{3k}$ (\SI{}{\GHz\: \um^3})}&
\textrm{$|\Delta|/2\pi$ (kHz) }&
\textrm{$R_c$ (\SI{}{\um})}\\ 
\colrule 
$\text{Rb}92s_{1/2}$\newline
$\text{K}95s_{1/2} $& $\text{Rb}92p_{1/2}$ \newline $ \text{K}94p_{1/2}$ &  -26.6 & 9 & 166 \\  

 $\text{Rb}96s_{1/2}$\newline 
 $\text{K}92s_{1/2} $ & $ \text{Rb}95p_{3/2}$ \newline $ \text{K}92p_{3/2}$ & -28.3 & -115 & 78 \\  
 
 $\text{Rb}97s_{1/2}$\newline 
 $\text{K}92s_{1/2} $ & $ \text{Rb}96p_{1/2}$ \newline $ \text{K}92p_{1/2}$ & -28.5 &-100  & 73 \\ 
 
 $\text{Rb}129s_{1/2}$ \newline
$\text{K}132s_{1/2}$ & $\text{Rb}129p_{3/2} $ \newline $ \text{K}131p_{3/2}$ &-103.2 & 53 &156\\ 

$\text{Rb}82d_{5/2}$ \newline
$\text{K}100d_{5/2}$ & $\text{Rb}83p_{3/2} $ \newline $ \text{K}102p_{3/2}$ &-52 & 290 & 59\\



  $\text{Rb}95s_{1/2}$ \newline
$\text{K}90d_{3/2}$ & $\text{Rb}95p_{1/2} $ \newline $ \text{K}91p_{1/2}$ & 43&  290 &55 \\

  $\text{Rb}100s_{1/2}$ \newline
$\text{K}100d_{5/2}$ & $\text{Rb}99p_{3/2} $ \newline $ \text{K}102p_{3/2}$ & 44.1&  90 &93 \\

  $\text{Rb}87d_{5/2}$ \newline
$\text{K}103s_{1/2}$ & $\text{Rb}86f_{5/2} $ \newline $ \text{K}102p_{3/2}$ & 10.0&  27 &85 \\
  $\text{Rb}88d_{3/2}$ \newline
$\text{K}104s_{1/2}$ & $\text{Rb}87f_{5/2} $ \newline $ \text{K}103p_{3/2}$ & -39 &  428 &57 \\
\end{tabular}
\end{ruledtabular}
\end{table*}
For initial pair states $\ket{n_{\rm Rb}s_{1/2}}\otimes \ket{n_{\rm K}s_{1/2}}$, the investigation can be restricted to pair states with angular momentum parts $\ket{p_{1/2}}\otimes \ket{p_{1/2}}$, $\ket{p_{1/2}}\otimes \ket{p_{3/2}}$, $\ket{p_{3/2}}\otimes \ket{p_{1/2}}$, or $\ket{p_{3/2}}\otimes \ket{p_{3/2}}$, leading to four possible $k$ channels. 
The principal quantum numbers of these dipole allowed states are, however, unrestricted, evoking a large number of coupled K-Rb pair state combinations. 
For all combinations of initial principal quantum numbers $50\le n_{\rm Rb},n_{\rm K}\le110$, we examine the variation of defect-minimized $C_{3k}$ coefficients and show the results on a grid in Fig.~\ref{figS_state}(a). We find the strongest $C_{3k}$ coefficients for $\Delta n = |n_\text{Rb} - n_\text{K}| \leq$ 5, which are visible as dark bands of near-diagonal elements in Fig.~\ref{figS_state}(a). The strongest resonances occur for combinations with small differences in the initial principal quantum number as previously reported for intraspecies systems, e.g., in Ref. \cite{Firstenberg2016}. Combined with their small F\"orster defects $|\Delta|/(2\pi) \lesssim$ \SI{12}{\MHz}, these states are promising candidates for large crossover distances, meaning a $C_\text{3}/R^3$ dependence over a wide range of interatomic separations, as follows from Eq.~(\ref{Cdistance}). This allows for stronger interactions at larger distances. In Fig.~\ref{figS_state}(b), we present the crossover distances corresponding to the pair combinations with the smallest F\"orster defects. The quasi-degenerate pair states with the longest crossover distance emerge as dark points, and Table~\ref{table1} lists the three combinations with largest $R_c$ \new{for initial $s+s$ states}. Additional resonances are listed in Appendix~\ref{C}. The state pair that maximizes $R_c$ [as indicated with an arrow in Fig.~\ref{figS_state}(b)] simultaneously displays our survey's minimum F\"orster defect of only \SI{9}{kHz}. To our knowledge this defect is orders of magnitudes smaller than defects reported for any other Rydberg system, and leads to a crossover distance in the \SI{100}{\mu m}-range, exceeding previously reported values of $\lesssim$ \SI{20}{\mu m} for $n <$ 90 in a Cs-Rb system \cite{Beterov2015} .  
\begin{figure*}
	\includegraphics[clip, trim= 0.5cm 1.5cm 0.5cm 1.5cm,width=0.91\textwidth]{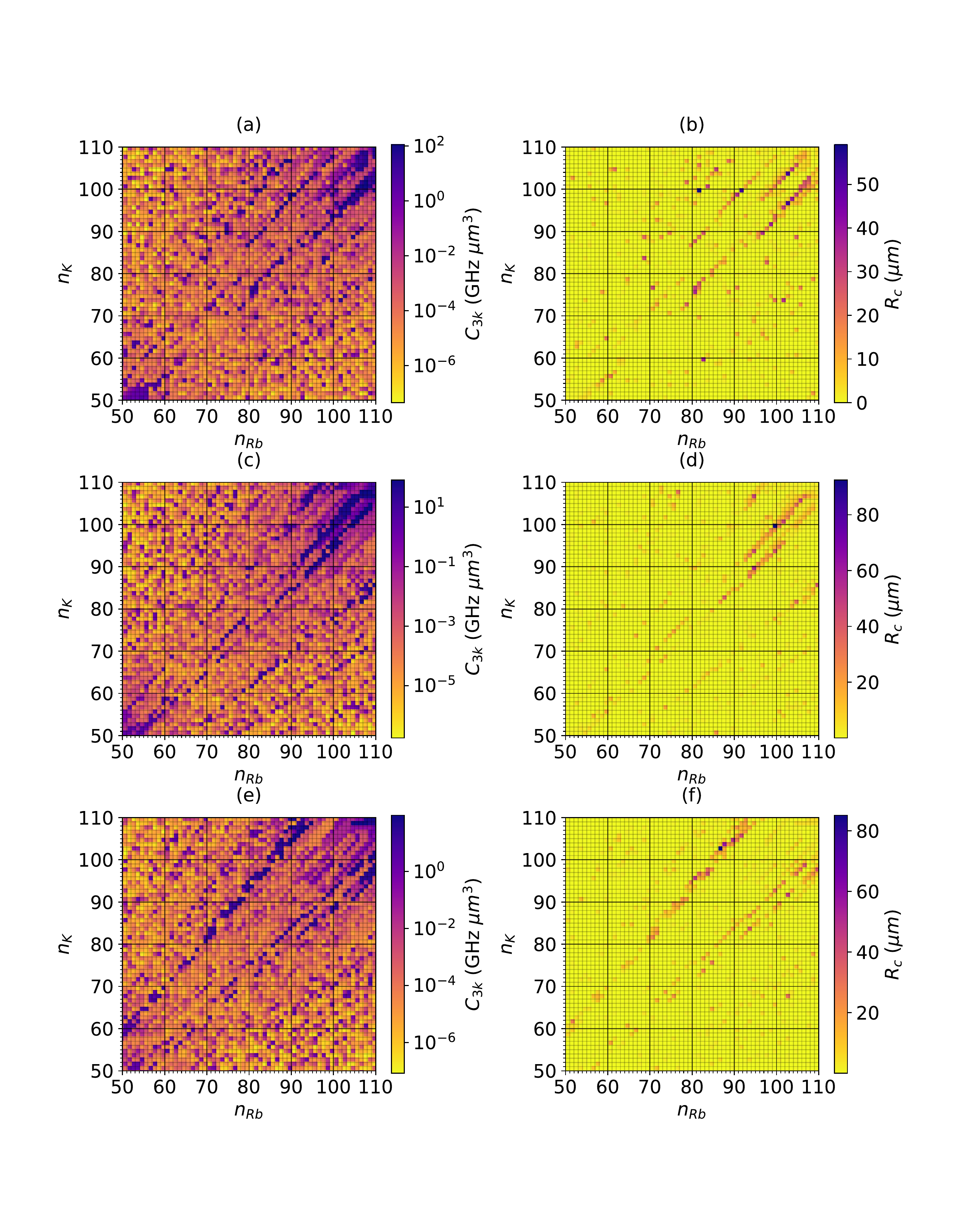}
	\caption{Calculated $|C_{3k}|$ coefficients and crossover distances $R_c$ for K-Rb pair states with the smallest F\"orster defects $|\Delta|$ for principal quantum numbers 50 $\leq n_\text{Rb}, n_\text{K} \leq$ 110 and different total angular momentum states. The initial pair states are $d + d$ for panels (a) and (b), $s + d$ for panels (c) and (d), and $d + s$ for panels (e) and (f). The combinations with the largest $R_c$ are listed in Table~\ref{table1} and in Table~\ref{table_s} in Appendix~\ref{C}.}
	\label{fig_d}
\end{figure*}
\subsection{$s+d, d+s$, and $ d+d$ channels}
In addition to $ns$ Rydberg states, $nd$ states can be excited by two-photon transitions. For experimental implementations, $d$ states are attractive because the dipole matrix elements for transitions from an intermediate $p_{3/2}$ state to a Rydberg $nd_{5/2}$ state can be twice as large as for the respective $ns_{1/2}$ states, enabling large Rabi frequencies for lower light powers. For interspecies systems, the excitation wavelengths are unique to each species, which necessitates four different excitation frequencies, independent of the choice of combinations of $s+s$, $d+d$ or $s+d$ state. For example, in our laboratory, EIT schemes are implemented using wavelengths of 480, and \SI{780}{\nm} for $^{87}\text{Rb}$ and 456 and \SI{767}{\nm} for $^{40}\text{K}$. As a result, each species can be addressed individually. 
\newline
We investigate initial pair states consisting of combinations of Rydberg $ns$ and $nd$ states. Compared to the $s+s$ pair states, the combination of $s$ and $d$ states leads to a larger number of allowed angular momentum channels, increasing the likelihood of channels with minimal interaction. Altogether, there are 12 possible $k$ channels for
$s+d \rightarrow p+p, p+f$, and 38 combinations for $d+d \rightarrow p+p, f+p, p+f, f+f$. As was done for the $s+s$ channels, we calculate interaction coefficients and crossover distances for the channels with the smallest F\"orster defects for the initial pair states $\ket{n_{\rm Rb}s_{1/2}}\otimes \ket{n_{\rm K}d_{j^{\text{K}}}}$, $\ket{n_{\rm Rb}d_{j^{\text{Rb}}}}\otimes \ket{n_{\rm K}s_{1/2}}$, and $\ket{n_{\rm Rb}d_{j^{\text{Rb}}}}\otimes \ket{n_{\rm K}d_{j^{\text{K}}}}$ and we present our findings in Fig.~\ref{fig_d}. Out of the 3$\times$3600 pair combinations with minimized F\"orster defect we find 15 state combinations with crossover distances $>\SI{40}{\mu m}$, mainly for larger principal quantum numbers of $n \approx 100$. Pair state combinations with $R_c$ exceeding $>\SI{55}{\mu m}$ are listed in {Table\,\ref{table1}}. For the $s+d$, $d+s$, and $d+d$ pair states, the largest $R_c$ values are found for $\Delta n = |n_\text{Rb} - n_\text{K}| \leq 8$, as, for instance, for  
\begin{align*}\new{
\ket{\text{Rb}100s_{1/2}} \otimes\ket{\text{K}100d_{5/2}} \leftrightarrow \ket{\text{Rb}99p_{3/2}}\otimes\ket{\text{K}102p_{3/2}},}
\end{align*}
with $C_{3k}$ = \SI{44.1}{\GHz\,\mu m^3} and $\Delta/(2\pi)$ = \SI{90}{\kHz}. 
\newline
In contrast to a single-species system, for two species the pair states are not energetically identical under interchange of the atomic quantum numbers, e.g. $E(\text{Rb}92s_{1/2}\,\text{K}95s_{1/2}) \neq E(\text{Rb}95s_{1/2}\,\text{K}92s_{1/2})$. As a result, the number of pair state energies is doubled and the probability to find pair state combination with small F\"orster defect $\Delta$ increases. \new{The quantum defects for K and Rb, which determine the pair state energies, obey $\delta_{nlj,\text{Rb}}- \delta_{nlj,\text{K}} \approx 1$ for $s,p,$ and $d$ states. Therefore, the pair state energies of $n$K-$n$Rb resemble those of $n$Rb$(n-1)$Rb \footnote{The energy of the Rydberg levels in the context of Quantum Defect Theory (QDT) is given by $E_{n,\text{Rb}} = - \frac{R_y}{(n-\delta_\text{Rb})^2}\approx - \frac{R_y}{(n-[\delta_\text{K}+1])^2} = -\frac{R_y}{([n-1]- \delta_\text{K})^2} = E_{n-1,\text{K}}$, where $R_y$ is the Rydberg constant and ${\delta_{nlj,\text{Rb}}- \delta_{nlj,\text{K}} \approx 1}$.}. We observe in Figs.~\ref{figS_state}(a) and ~\ref{fig_d}(a) dark bands with large $C_{3k}$ on both sides of the diagonal. Unlike for a single-species system \cite{Paris-Mandoki2016} we do not expect a full reflection symmetry about the diagonal for the $C_{3k}$ coefficients. However, due to our preselection of F\"orster defect minimized pair state combinations and the similarity of the K and Rb quantum defects, a pattern similar to the single-species Rb-Rb scenario occurs.} 
Large interaction coefficients cannot solely be found for small differences $\Delta n = |n_1 -n_2| \approx 5$ in the initial $n$ numbers of the two atoms. For the $d+s$ combination, the largest interaction coefficients and crossover distances appear for $|n_\text{Rb} - n_\text{K}|= $ 13, namely for 
\begin{align*}
\ket{\text{Rb}87d_{5/2}\text{K}100s_{1/2}} \leftrightarrow \ket{\text{Rb}86f_{5/2}\text{K}102p_{3/2}},
\end{align*}
with $C_{3k}$ = \SI{10}{\GHz\, \mu m^3} and $\Delta/(2\pi)$ = \SI{27}{\kHz}. \newline
 
The variety of possible $k$ channels for each combination of $n_\text{Rb}$ and $n_\text{K}$ allows for a large number of states with F\"orster defect $|\Delta|/(2\pi) <$ \SI{12}{\MHz} and interaction strengths $>$ \SI{1}{\GHz\,\mu m^3}, resulting in crossover distances $>$ \SI{8}{\mu m}, which we list in the Appendix~\ref{C} in Table~\ref{table_s}. This abundance of state combinations makes it possible to access F\"orster resonances at nearly zero electric field for a variety of Rydberg states.

\subsection{Comparison with different Rydberg-species combinations}
\label{sec3C}
In our identification of experimentally accessible pair states with long-range interaction, we used the F\"orster defect as a first criterion to select K-Rb states out of the vastness of possible state combinations. The subset of pair states with both small $\Delta$ and large interaction coefficient $C_{3k}$ allowed us to find pair states with large crossover distances in the absence of external fields. For comparison we \new{calculate and} list in Table~\ref{table_comp} the pair states with the smallest $\Delta$ for single-species and two-species combinations of Cs, Rb and \new{K --- }the most frequently used alkali Rydberg species. We restrict our examination to initial pair states $\ket{n_{a}s_{1/2}}\otimes \ket{n_{b}s_{1/2}}$ where $n <$ 100. For Cs-Cs, Rb-Cs and Rb-Rb pair states we find values in agreement with the calculations in Ref. \cite{Beterov2015}. For Rb-Rb, K-K, Cs-Cs, and Cs-K, the smallest F\"orster defects are in the order of a few hundred \SI{}{\kHz}. By contrast, for the K-Rb system, the pair state $\ket{\text{Rb}92s_{1/2}}\otimes\ket{\text{K}95s_{1/2}} \leftrightarrow \ket{\text{Rb}92p_{1/2}}\otimes\ket{\text{K}94p_{1/2}}$, which
we singled out in Sec.~\ref{sec2a}, stands out with its ultrasmall F\"orster defect of a mere \SI{9}{\kHz} and a comparatively large interaction strength of $C_\text{3k}$ = \SI{26.6}{\GHz\: \um^3}. Interestingly, the K-Rb system gives us more choices of pair states with small F\"orster defects in comparison to the other species combinations of Table~\ref{table_comp} \footnote{More K-Rb pair state combinations can be found in Table~\ref{table1} and Table~\ref{table_s} in Appendix~\ref{C}.}. Table~\ref{table_comp} also lists the pair states with the largest crossover distance $R_c$ for each species combination, where $\ket{\text{Rb}92s_{1/2}}\otimes\ket{\text{K}95s_{1/2}}$ reappears with a maximum of \SI{166}{\mu m}, much larger than values of 20 to \SI{40}{\mu m} for all other combinations. While in a practical situation the crossover distance will depend on the spin-dependent angular factor $\mathcal{S}_{k}$ [Eq.~(\ref{ceq2})], and we are only calculating the upper limit for $R_c$, the above discussion promotes K-Rb as a promising system for zero-field F\"orster resonances. 
\newline
\newline
\new{An alternative} approach for the classification of interacting pair states employs $C_{3k}$ coefficients as a first criterion. Such an examination was performed for Rb-Rb pair states in Ref. \cite{Paris-Mandoki2016} in $n_1s$ and $n_2d$ Rydberg fine structure states with $30\leq n_1,n_2 \leq 100$, and $|\Delta|/(2\pi) \leq$ \SI{500}{MHz}. These criteria are similar to the ones used in Secs.~\ref{sec3}A and B. In contrast to a two-species system, for two identical Rb atoms prepared in a pair state $\ket{n_1 j_1}\otimes\ket{n_2 j_2}$ a resonant coupling to the pair state $\ket{n_2 j_2}\otimes\ket{n_1 j_1}$ with $\Delta$ = 0 will always exist. 
The interaction gives rise to coherent exchange of the internal states of the atoms \cite{Browaeys2020}, which can be observed as an oscillation (also called flip-flop or hopping) between the states $\ket{n_1 j_1}\otimes\ket{n_2 j_2}$ and $\ket{n_2 j_2}\otimes\ket{n_1 j_1}$. Hence, for single-species systems with two atoms in different initial states it is necessary to consider two channels with identical F\"orster defect $\Delta$, but generally different coefficients $C_{3k}$ and $C_{3k}^{'}$, as we show for an exemplary initial state $\ket{n_1s_{1/2}}\otimes\ket{n_2d_{5/2}}$ that couples to $\ket{n_3p_{3/2}}\otimes\ket{n_4p_{3/2}}$ as
\begin{widetext}
\begin{equation}\label{hopping}
  \underbrace{\ket{n_1 s_{1/2}} \otimes \ket{n_2 d_{5/2}}}_{
    \tikz[remember picture,overlay] \coordinate (A);
  }
  \xleftrightarrow[|\Delta|]{C_{3k}}
  \ket{n_3 p_{3/2}} \otimes \ket{n_4 p_{3/2}}
  \xleftrightarrow[|\Delta|]{C_{3k}'}
  \underbrace{\ket{n_2 d_{5/2}} \otimes \ket{n_1 s_{1/2}}}_{
    \tikz[remember picture,overlay]
      \draw[<->] (0,0) -- (0,-2ex) -- node[below] {hopping} ([yshift=-2ex]A) -- (A);
  }.
  \vspace{0.5cm} 
\end{equation}
\end{widetext}
The magnitude of the interaction coefficients $C_{3k}$ and $C_{3k}^{'}$ is ultimately determined by the radial wave function overlap between the Rydberg states. The peak position of the radial probability density and therefore the radial wave functions scale with $n^{2}$ \cite{Pritchard_2012}. Consequently $C_{3k}$ and $C_{3k}^{'}$ rapidly decrease as $\Delta n$  grows and interactions are dominated by a small number of close-lying $n$ states with respect to the initial principal quantum numbers \cite{Beterov2015, Paris-Mandoki2016}. Returning to the example of Eq.~(\ref{hopping}), for $C_{3k}$ this relates to changes in $n$ of $\Delta n_1 =|n_1 - n_3|$ and $\Delta n_2 =|n_2 - n_4|$ for $C_{3k}$ , while $C_{3k}^{'}$ connects to changes $\Delta n_1^{'} =|n_3 - n_2|$ and $\Delta n_2^{'} =|n_4 - n_1|$. 
If the magnitudes of $C_{3k}$ and $ C_{3k}^{'}$, are similar and the F\"orster defect $\Delta$ of the pair states is small, the two atoms can exchange their internal states as a consequence of the dipole-dipole interaction \cite{Paris-Mandoki2016} via the hopping process indicated in Eq.~(\ref{hopping}). Such resonant excitation exchange with Rydberg atoms opens up the possibility to implement spin-exchange Hamiltonians, which are useful for quantum simulators, as well as for 
the study of quantum magnetism, and transport phenomena \cite{Browaeys2020,Barredo_2015, Whitlock_2017}. The exchange interaction, however, is inherently resonant and as a result cannot be turned on and off with an external field, as it is possible for F\"orster resonances. For blockade experiments the main interest lies in a shift of energy due to the dipole-dipole interaction and state combinations with suppressed hopping dynamics. This is achieved if one of $C_{3k}$  or $C_{3k}^{'}$ is $\approx$ 0 while the other takes on a large value, which can be realized by choosing $\Delta n_1, \Delta n_2 \ll \Delta n_1^{'},\Delta n_2^{'}$ (or alternatively $\Delta n_1^{'},\Delta n_2^{'} \ll \Delta n_1, \Delta n_2$). Altogether, this greatly restricts the choice of F\"orster resonances in single-species systems.
For two-species systems ($Z_1 \neq Z_2$), as K-Rb, we are not limited to such states, as for these the F\"orster defects for $\ket{Z_1n_1 l_1 j_1}\otimes\ket{Z_2n_2 l_2j_2} \leftrightarrow \ket{Z_1n_3l_3 j_3}\otimes\ket{Z_2n_4 l_4j_4}$ and $\ket{Z_1n_1 l_1j_1}\otimes\ket{Z_2n_2 l_2j_2} \leftrightarrow \ket{Z_1n_4 l_4j_4}\otimes\ket{Z_2n_3 l_3j_3}$ are generally of very different size, and a hopping process cannot take place. Relating this to a point ($n_\text{1,Rb}n_\text{2,K}$) on the grids of Figs.~\ref{figS_state} and \ref{fig_d}, this means we do not have to take into consideration the interaction connected to ($n_\text{2,Rb}n_\text{1,K}$), as it would be the case for a single-species system. 
In summary, we find $C_{3k}$ coefficients of similar strength to other species combinations, as for $\ket{\text{Rb}92s_{1/2}}\otimes\ket{\text{K}95s_{1/2}} \leftrightarrow \ket{\text{Rb}92p_{1/2}}\otimes\ket{\text{K}94p_{1/2}}$ we obtain a value of $C_{3k}$ = \SI{26.2}{\GHz\: \um^3}, which is comparable to the strongest $C_{3k}$ for a zero-field Rb-Rb resonance in Ref. \cite{Paris-Mandoki2016}. However, for the K-Rb pair state the F\"orster defect is only $2\pi \times$\SI{9}{\kHz} compared to $2\pi \times$\SI{3.5}{\MHz} of the latter. This makes the K-Rb pair state to an outstanding candidate for long-range interaction at zero-field. 
\begin{table*}[t]
\begin{ruledtabular}
\caption{Comparison of the smallest F\"orster defects $|\Delta|$ and largest crossover distances $R_c$ for single and two-species pair states composed of the alkali Rydberg atoms K, Rb and Cs initially in the $s_{1/2}$ state and for $n < 100$. \new{The values for all presented Rydberg species are calculated with our code and we find agreeing values for Rb-Rb, Rb-Cs and Cs-Cs with \cite{Beterov2015}. Additionally, in Table~\ref{table_KK} in Appendix~\ref{C} we list F\"orster defects and interaction coefficients for the single-species K system for pair states with strong interaction and $n<100$.}}
\begin{tabular}{cccrr}
smallest $|\Delta|$\\
Species & Initial s-State & Final State & $|\Delta|$/(2$\pi$) (kHz)& $C_{3k}$ (\SI{}{\GHz\: \um^3})\\ \hline
Rb-Rb & $89,64$&$90p_{3/2}\,63p_{3/2}$ & 694 & 1.4 \\
K-K & $59,62$&$59p_{3/2}\,61p_{1/2}$ & 274 &4.6 \\
Cs-Cs & $98,71$&$99p_{1/2}\,70p_{3/2}$ & 324  & -1.8 \\
Rb-Cs & $92,63$&$90p_{3/2}\,63p_{1/2}$ & 1828 & -1.5 \\
Rb-K & $92,95$ & $92p_{1/2}\,94p_{1/2}$ & 9 & -26.6 \\ 
Cs-K &  $95,90$ & $94p_{3/2}\,90p_{1/2}$ & 656 & 26.1 \\ 
\\
largest $R_c$\\
Species & Initial s-State & Final State & $R_c$ (\SI{}{\um})& $C_{3k}$ (\SI{}{\GHz\: \um^3})\\ \hline
Rb-Rb & $84,81$ &$83p_{1/2}\,81p_{1/2}$ & 19.3  & -15.7\\
K-K & $82,78$&$81p_{1/2}\,78p_{1/2}$ & 34.4 & -14.5 \\
Cs-Cs & $98,71$&$99p_{1/2}\,70p_{3/2}$ & 20.1 & -1.8  \\
Rb-Cs & $99,96$&$98p_{3/2}\,96p_{1/2}$ & 26.2 & 32.1  \\
Rb-K & $92,95$ & $92p_{1/2}\,94p_{1/2}$ &  166.4 & -26.6 \\
Cs-K &  $95,90$ & $94p_{3/2}\,90p_{1/2}$ & 38.3 & 26.1 \\ 
\end{tabular}
\label{table_comp}
\end{ruledtabular}
\end{table*}
\section{PAIR POTENTIALS, EXTERNAL FIELDS AND ANGULAR DEPENDENCE FOR $\ket{\text{Rb}92s_{1/2}}\otimes\ket{\text{K}95s_{1/2}} \leftrightarrow \ket{\text{Rb}92p_{1/2}}\otimes \ket{\text{K}94p_{1/2}}$}
\label{sec4}
\begin{figure}[t]
\includegraphics[clip, trim= 0.1cm 3.1cm 0.2cm 0.4cm,width=0.98\columnwidth]{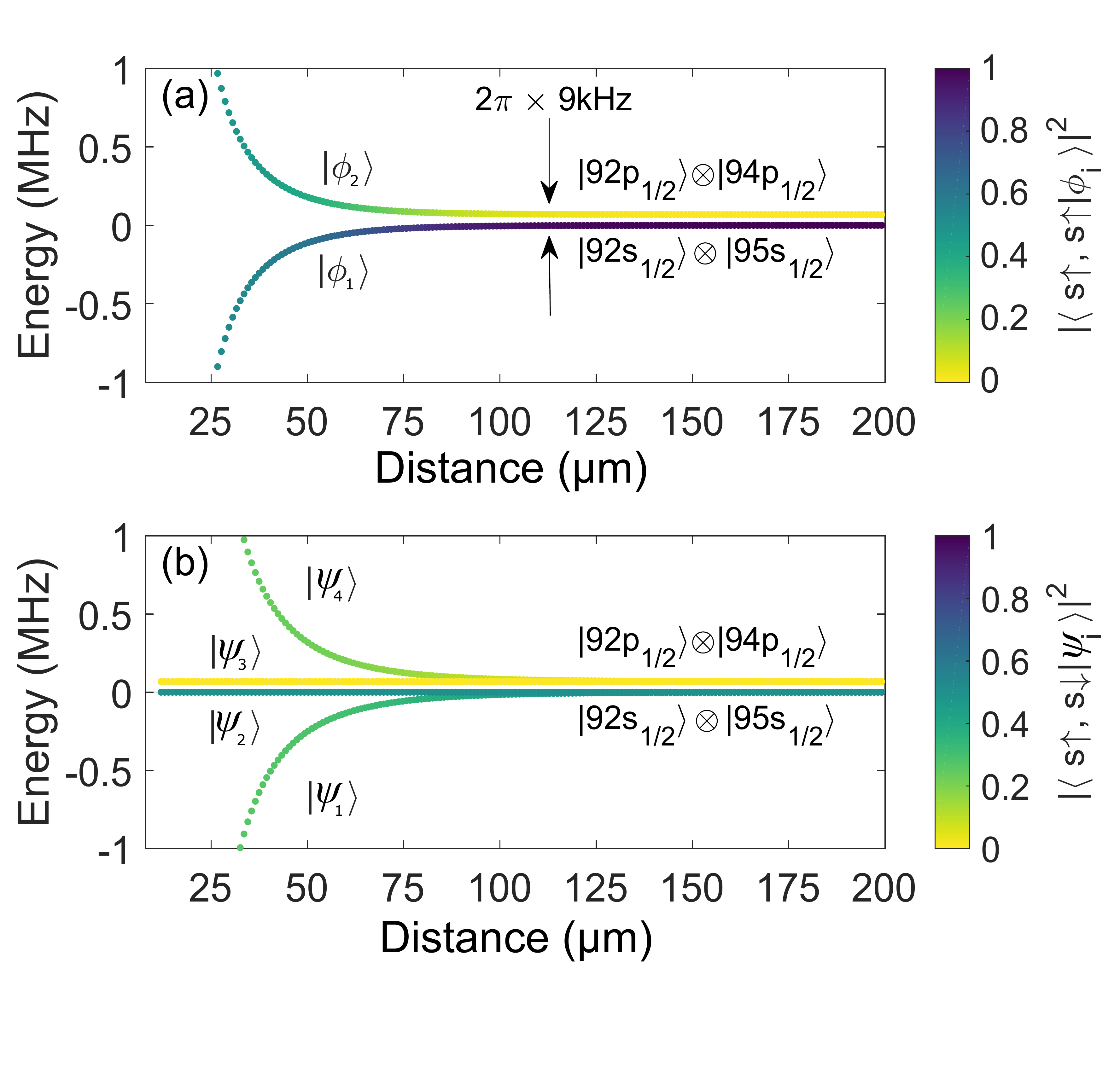}
\caption{\label{figZeeman} Zero-field K-Rb pair potentials in the vicinity of the pair states $\ket{\text{Rb} 92s_{1/2}m_{j\text{Rb}}} \otimes \ket{\text{K} 95s_{1/2}m_{j\text{K}}}$ for initial spin states (a) $(m_{j\text{Rb}},m_{j\text{K}})$ = $(\uparrow,\uparrow)$ and (b) ${(m_{j\text{Rb}},m_{j\text{K}}) = (\uparrow,\downarrow)}$. Only the coupling $\ket{s_{1/2}} \otimes \ket{s_{1/2}} \leftrightarrow \ket{p_{1/2}} \otimes \ket{p_{1/2}}$ (incorporating $m_j$) is included in the calculation. The color code illustrates the overlap $|\braket{s\uparrow, s\uparrow | \phi_i}|^2$ in panel (a) and $|\braket{s\uparrow, s\downarrow | \phi_i}|^2$ in panel (b) between the noninteracting eigenstate $\ket{\alpha_a\alpha_b}$ for $V_\text{dd} = 0$ and the eigenstates $\ket{\phi_i}$ and $\ket{\psi_i}$ of the interacting system. }
\end{figure}
\noindent
So far, our discussion has been based on F\"orster defects and the spin-independent interaction coefficients $C_{3k}$ and has not taken into consideration the angular factor $\mathcal{S}_{k}(\theta)$ \new{appearing in Eq.~(\ref{ceq2}),} which depends on the magnetic quantum numbers, the system geometry, and externally applied electric and magnetic fields. In the following, we extend our treatment to include the magnetic quantum numbers $m_j$ in the description and we account for Zeeman degeneracy. \newline
To gain insights into the angular dependence of the interaction, we focus on the coupling of the pair state $\ket{\text{Rb}92s_{1/2}}\otimes \ket{\text{K}95s_{1/2}} $ to $\ket{\text{Rb}92p_{1/2}}\otimes \ket{\text{K}94p_{1/2}}$ that emerged from our search for candidates with long-range interaction in Sec.~\ref{sec3}. The two pair states have a F\"orster defect of $\Delta/(2\pi)$ = \SI{9}{kHz}. At the same time, the resonance is --- with a separation of more than \SI{20}{\MHz} --- well isolated from even the closest fine structure channels $\ket{\text{Rb}92p_{1/2}}\otimes \ket{\text{K}94p_{3/2}}$, $\ket{\text{Rb}92p_{3/2}}\otimes \ket{\text{K}94p_{1/2}}$ and $\ket{\text{Rb}92p_{3/2}}\otimes \ket{\text{K}94p_{3/2}}$, as presented in Table~\ref{table2}. For brevity, we represent the spin-dependent pair states $\ket{\text{Rb}92s_{1/2}m_{j_{\alpha{_a}}}}\otimes \ket{\text{K}95s_{1/2}m_{j_{\alpha{_b}}}}$ and $\ket{\text{Rb}92p_{1/2}m_{j_{\beta{_a}}}}\otimes \ket{\text{K}94p_{1/2}m_{j_{\beta{_b}}}}$ as $\ket{s\, m_{j_{\alpha{_a}}}, s\, m_{j_{\alpha{_b}}}}$ and $\ket{p\, m_{j_{\beta{_a}}}, p\, m_{j_{\beta{_b}}}}$, where we denote $m_{j} = 1/2$ as $\uparrow$ and $m_{j} = -1/2$ as $\downarrow$. \newline
\begin{table}[b]
\caption{\label{table2}%
F\"orster defects and $C_{3k}$ coefficients for the four fine structure $k$-channels of $\ket{n_1s_{1/2}}\otimes\ket{n_2s_{1/2}} \leftrightarrow \ket{n_1p_j}\otimes\ket{(n_2-1)p_{j'}} $ for 92Rb-92Rb, 95K-95K and 92Rb-95K. For the single-species systems the $C_{3k}$ and $\Delta$ are of similar magnitude for all $k$, which indicates isotropic interaction characteristics. For the K-Rb case the supremacy of the F\"orster defect of a single channel leads to an overall stronger interaction.
}
\begin{tabular}{ccccc}
\hline \hline
  \noalign{\smallskip} \noalign{\smallskip}

$(j,j')$ & ${(1/2,1/2)}$ &${(3/2,1/2)}$ &${(1/2,3/2)}$& ${(3/2,3/2)}$ \\  \hline 
$|\Delta|/2 \pi$ (MHz)&&&& \\ 
\noalign{\smallskip}

Rb-Rb & 524& 398 & 	402		  & 276\\
K-K  &622		& 596& 599 	& 571\\
Rb-K  	&0.009 & 121&  26 & 147\\
  \noalign{\smallskip}
 $C_{3k}$  (\SI{}{GHz\, \mu m^3})&&&& \\ \noalign{\smallskip}
Rb-Rb & -24.7 & 24.3 & 25.2	& -24.8  \\
K-K & -29.4 		& 29.3  & 29.5	& -29.4 \\
Rb-K  & 26.2 & -26.7  & -26.3& -26.6  \\
 \noalign{\smallskip}\hline \hline
\end{tabular}
\end{table}
\subsection{Zeeman degeneracy at $\theta$ = 0}
\label{sec4a}
For $\theta = 0$, the quantization axis is directed along the interatomic axis of the two atoms and the total spin projection along this axis, $M = {m_{\alpha_{a}}} {+}{ m_{\alpha_{b}}}$, is conserved as a direct consequence of the rotation invariance of the Hamiltonian. The dipole-dipole interaction therefore only couples to pair states with $M'=M$, where $M'= {m_{\beta_{a}}}{+}{m_{\beta_{b}}}$. Hence, a state $\ket{s \uparrow,s\uparrow}$ solely couples to the pair state $\ket{p \uparrow, p\uparrow}$, satisfying the selection rule $\Delta M = M' -M = 0$. At a given atomic separation $R$, the two eigenstates $\ket{\phi_1}$ and $\ket{\phi_2}$ of the coupled system are superpositions of the eigenstates of the Hamiltonian in Eq.~(\ref{ham}) for zero interaction ($V_\text{dd}=0$), namely $\ket{s\uparrow, s\uparrow}$ and $\ket{p \uparrow ,p \uparrow}$. It follows that $\ket{\phi_i} = c_{i,s}(R) \ket{s\uparrow, s\uparrow} + c_{i,p}(R) \ket{p\uparrow ,p\uparrow}$, with the normalization $|c_{i,s}|^2 +|c_{i,p}|^2 =1$.  %
\newline
In Fig.~\ref{figZeeman}(a) we show the pair potential for the coupling of $\ket{\text{Rb}92s_{1/2}\uparrow}\otimes \ket{\text{K}95s_{1/2}\uparrow}$ and $\ket{\text{Rb}92p_{1/2}\uparrow}\otimes \ket{\text{K}94p_{1/2}\uparrow}$, assuming a two-level approximation. For large interatomic separations, the coupling between the two pair states is negligible, and they are separated in energy by their F\"orster defect $\Delta/(2\pi)$ = \SI{9}{\kHz}, as shown in Fig.~\ref{figZeeman}(a).
As the interaction increases with decreasing separation, the two energy level repel and the corresponding eigenstates become mixtures of $\ket{s \uparrow,s\uparrow}$ and $\ket{p \uparrow, p\uparrow}$. The levels in Fig.~\ref{figZeeman}(a) are color coded according to their $\ket{\s \uparrow, s\uparrow}$ admixture, visualizing how the lower level $\ket{\phi_1}$ connects adiabatically to $\ket{\s \uparrow, s\uparrow}$. 
Hence, for smaller distances $\ket{s \uparrow,s\uparrow}$ and $\ket{p \uparrow,p\uparrow}$ are no longer eigenstates of the coupled system, and dependent on the state preparation the system will oscillate between them \cite{Paris-Mandoki2016}. For even smaller distances ($R \lesssim$ \SI{25}{\um}), the 1/$R^3$ dependence of the dipole-dipole interaction dominates and an increased number of states with larger F\"orster defects contribute to the scenario. The two-level approximation is not valid anymore, and Eqs.~(\ref{twolevel1}) and (\ref{Ufull}) are no longer applicable. The rather chaotic energy-level diagram in this ``spaghetti region" \cite{PhysRevA.53.1349} can only be determined numerically and goes beyond the scope of our investigations in Fig.~\ref{figZeeman}. 
\newline
In Fig.~\ref{figZeeman}(b), we show the pair potential associated with the antiparallel state $\ket{s \uparrow, s\downarrow}$, for which $M=0$. At large atomic separation, this state can be decomposed into equal singlet $\ket{\psi^{(s)}_S}$ and triplet $\ket{\psi^{(s)}_T}$ (with $M = 0$) parts as $\ket{s \uparrow, s\downarrow} = \frac{1}{\sqrt{2}}[\ket{\psi^{(s)}_S}+\ket{\psi^{(s)}_T}]$. Upon decreasing the separation $R$, the states $\ket{\psi_1}$ and $\ket{\psi_2}$  that adiabatically connect to $\ket{\psi^{(s)}_T}$ and $\ket{\psi^{(s)}_S}$, respectively, are split in energy as a result of interaction between $\ket{\psi^{(s)}_T}$ and $\ket{\psi^{(p)}_T}=\frac{1}{\sqrt{2}}(\ket{p \uparrow,p\downarrow}+\ket{p \downarrow,p\uparrow})$, which causes $\ket{\psi_1}$ to shift.
As a consequence of the interaction, the energy level that adiabatically connects with $\ket{\psi^{(p)}_T}$ at large distances,  $\ket{\psi_4}$, is also shifted. Meanwhile, the states $\ket{\psi_2}$ and $\ket{\psi_3}$ that connect to (and in fact remain equal to) $\ket{\psi^{(s)}_S}$ and $\ket{\psi^{(p)}_S}=\frac{1}{\sqrt{2}}(\ket{p \uparrow,p\downarrow}-\ket{p \downarrow,p\uparrow})$, respectively, do not interact because they are the zero eigenvectors of the $\mathcal{S}_{k}$ matrix (for $M$ = 0) with eigenvalue 0 \cite{Walker_Saffman}. This results in a flat potential curve for $\psi_2$. Such channels are known as F\"orster zeros \cite{Zeros_Rydberg} and pose a limit for experiments that require large interaction strength for all possible angular momentum channels, often desired in the context of fully blockaded mesoscopic atomic ensembles \cite{Zeros_Rydberg}. 
The efficiency of the blockade is predominantly determined by the energy shift of the desired pair state, e.g. $\ket{s\uparrow, s\downarrow}$ in Fig.~\ref{figZeeman}(b). Because of the substantial overlap with the non-shifted singlet state, $|\braket{s\uparrow s\downarrow| \psi_2}|^2 = 0.5$ for all $R$, the dipole blockade is suppressed. \newline \newline
Generally, the polarization of the excitation light and the angular distribution of the atoms can play an important role and lead to nonzero overlap between the unperturbed eigenstate ($V_\text{dd} = 0$) and components of the coupled system with ${M = 0, \pm 1}$, some of which can have interaction coefficients $C_{3}(\theta) = C_{3k} \mathcal{S}_k(\theta)^{1/2}$ that are small or even zero due to the dependence of $\mathcal{S}_k(\theta)$ on $M$ and $\theta$. F\"orster zeros mainly emerge for $M$ = 0 channels \cite{Zeros_Rydberg}.
For the presented case one can excite the $M =1$ state, $\ket{s\uparrow, s\uparrow}$, and avoid components in zero interaction channels. The shift in energy, however, is smaller than for the interacting component of the $M =0$ channel in Fig.~\ref{figZeeman}(b). The crossover distances is reduced to $R_c$ = \SI{67}{\mu m}, whereas for $M = 0$ the largest $R_c$ of \SI{166}{\mu m} can be obtained with the drawback of a possible F\"orster zero.

\begin{figure}[t] 
\includegraphics[clip, trim= 0.0cm 0cm 0.0cm 0cm,width=0.98\columnwidth]{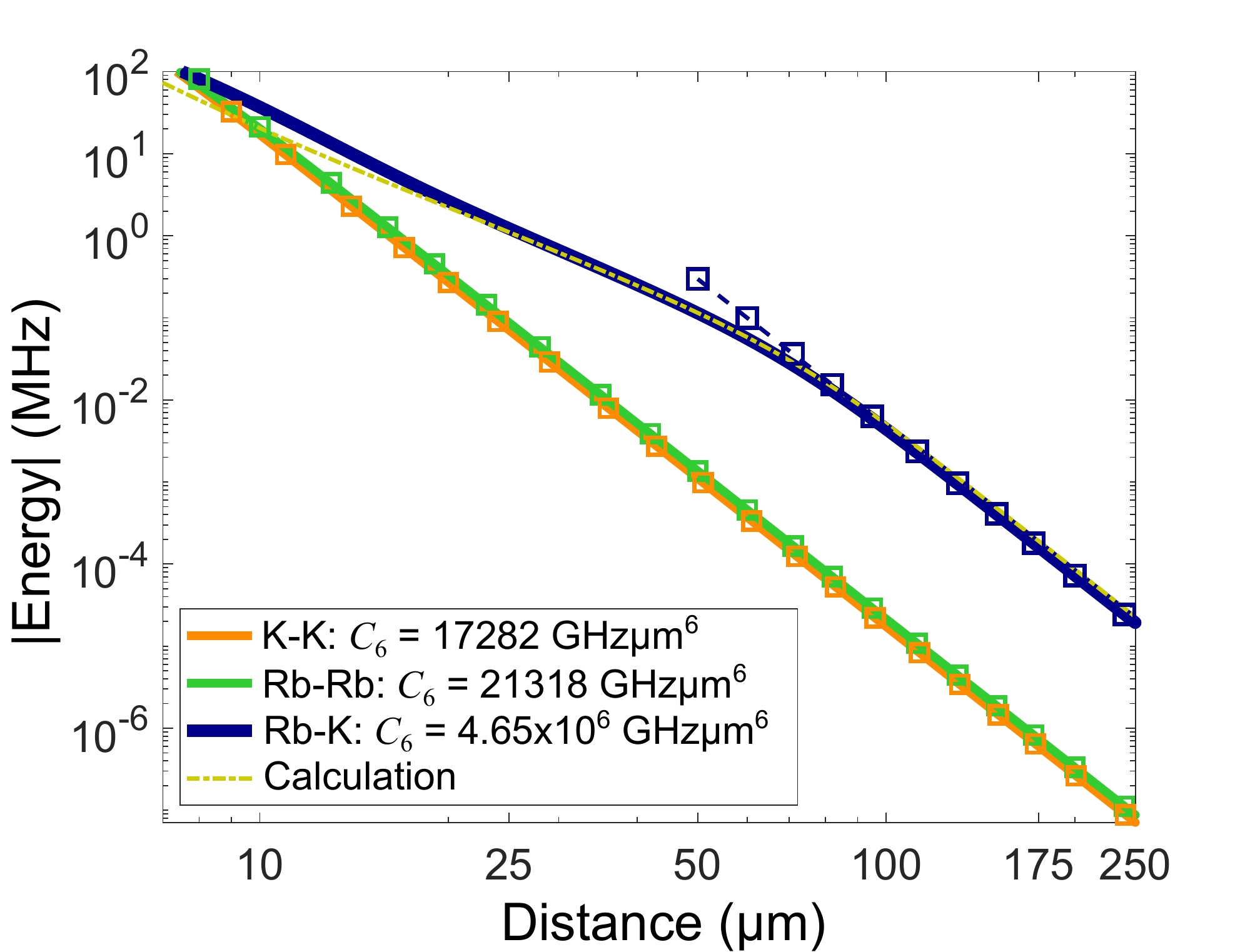}%
\caption{\label{figpot} Numerically calculated pair potentials dependent on the interatomic distance using Ref.~\cite{Weber2017} for $\ket{\text{Rb} 92s_{1/2} \uparrow\text{Rb} 92s_{1/2}\uparrow}$, $\ket{\text{K} 95s_{1/2}\uparrow\text{K} 95s_{1/2}\uparrow}$ and $\ket{\text{Rb} 92s_{1/2}\uparrow\text{K} 95s_{1/2}\uparrow}$ at $\theta$ = \SI{0}{\degree} and in the absence of external fields. $C_6$ coefficients are calculated pertubatively with Ref.~\cite{Durham_tutorial} and $C_6/R^6$ is represented by squares. The dashed yellow line is calculated by diagonalizing the Hamiltonian including the two energetically closest pair states.
}\end{figure}
\begin{figure*}[t]
\centering
\includegraphics[clip, trim= 0.0cm 0.0cm 0.0cm 0.0cm,width=1.98\columnwidth]{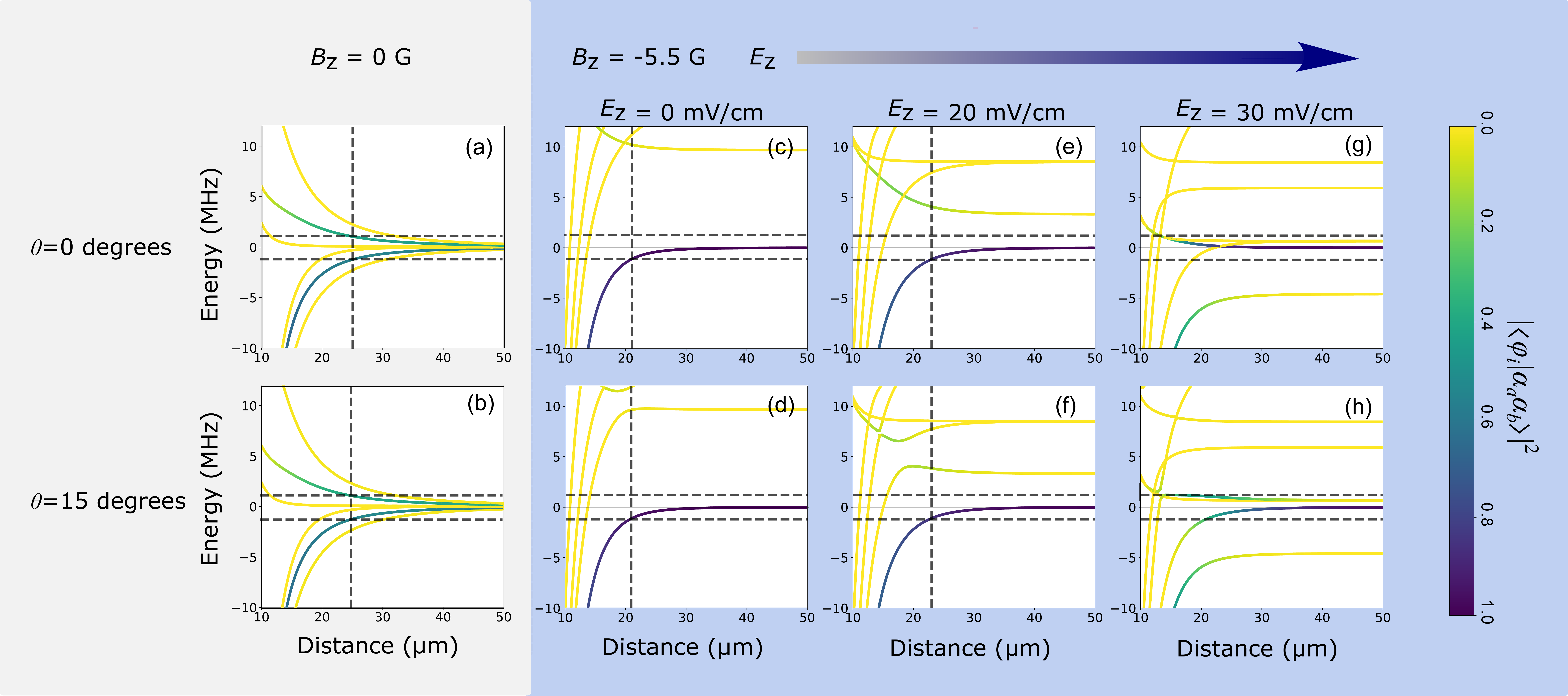}
\caption{\label{figZeeman2} Pair potentials in the vicinity of $\ket{\alpha_a\alpha_b} = \ket{\text{Rb}92s_{1/2}\uparrow}\otimes \ket{\text{K}95s_{1/2}\uparrow}$ for an increasing static electric field (left to right) and two angles $\theta$ = \SI{0}{\degree} and $\theta$ = \SI{15}{\degree}. A static magnetic field of \SI{-5.5}{G} (blue shaded background) lifts the Zeeman degeneracy in panels (c)--(h). The color code depicts the overlap with $\ket{\alpha_a,\alpha_b}$. For $E$ = \SI{0}{mV/cm} [(c), (d)] and $E$ = \SI{20}{mV/cm} [(e), (f)], $\ket{\alpha_a\alpha_b}$ does not mix with other states. For $E$ = \SI{20}{mV/cm}, the near-isotropic potential of Fig.~\ref{figZeeman3} can be obtained. The energy shift is $>$ \SI{1.2}{\MHz} (dashed lines) for R $<$ \SI{22}{\mu m}. A change to $E$ = \SI{30}{mV/cm} significantly changes the scenario [(g),(h)]. The pair states strongly mix, allowing for unwanted excitation of other states by a laser with an arbitrary chosen linewidth of \SI{1.2}{\MHz}.}
\end{figure*}
\noindent
For scenarios involving ensemble of atoms, it is important to not only understand the interaction between atoms in the target states $\ket{\text{Rb}92s_{1/2}\uparrow}\otimes \ket{\text{K}95s_{1/2}\uparrow} $, but also the interplay between the single-species states $\ket{\text{Rb}92s_{1/2}}\otimes\ket{\text{Rb}92s_{1/2}}$ and $\ket{\text{K}95s_{1/2}}\otimes\ket{\text{K}95s_{1/2}} $. In Fig.~\ref{figpot} we show numerically calculated pair potentials for K-Rb, Rb-Rb, and K-K for initial states $\ket{s\uparrow,s\uparrow}$ at $\theta$ = \SI{0}{}, and in the absence of external fields. 
For the single-species pair states the F\"orster defects to adjacent pair states are large (Table~\ref{table2}) compared to the interaction energies, and the pair potentials follow a power law scaling of $C_6({\theta}{=}{0})/R^6$, where the vdW coefficient $C_6(\theta)$ depends on the atomic energy levels and dipole matrix elements containing the angular momentum properties and can be calculated by second-order perturbation theory.
\newline
In contrast to the single-species potentials, the K-Rb potential shows a transition from the vdW regime to the resonant dipole-dipole regime at distances as large as ${\sim \SI{70}{\mu m}}$, which is in agreement to our calculated value of \SI{67}{\mu m} for $\ket{\uparrow,\uparrow}$ using a two-level approximation. The yellow line in Fig.~\ref{figpot} represents the results of the diagonalization under consideration of the two K-Rb pair states that interact the strongest with $\ket{\text{Rb}92s_{1/2}\uparrow}\otimes \ket{\text{K}95s_{1/2}\uparrow}$. For $R >$ \SI{25}{\mu m} the description is valid, whereas for shorter distances more states have to be taken into account. With a $C_6$ coefficient that is approximately 200 times larger than for the single-species systems, the interaction is distinctively stronger \new{for K-Rb}. This can be attributed to the up to three orders of magnitude smaller $\Delta$ for one of the four fine-structure channels, listed in Table~\ref{table2}. 
A maximized interaction between the two-species pair state in combination with a minimized interaction between the single-species states can have interesting applications, as we show in Sec.~\ref{sec5}. 

\subsection{Tuning with external \textit{E} and \textit{B} fields}
We finally examine a more general case where the quantization axis is defined by a coaxial external electric and magnetic field and lies at arbitrary angles $\theta$ with respect to the interatomic axis. This can be seen as a rotation of the interatomic axis relative to the fixed laboratory frame defining the quantization axis. For the zero field case, as considered in Sec.~\ref{sec4a}, the pair states are Zeeman degenerate. The degeneracy can be lifted by applying a small external magnetic field of a few Gauss, which causes spin-dependent shifts in energy. \new{If laser light with suitable polarization is chosen \cite{Walker_Saffman}, addressing a single Zeeman pair state can be simplified and undesirable F\"orster zero states can be avoided.} Because of the high density of pair states, state-mixing caused by dipole-dipole interactions can frequently occur and as such coupling to a single well-isolated channel can only be obtained in special geometries. Even when such configurations can be realized, external electric fields can Stark-shift magnetic substates into resonance causing population of unwanted channels and a breakdown of the Rydberg blockade, as pointed out in Ref. \cite{Browaeys_2018b}.\newline\noindent%
\begin{figure}[b]
\centering 
\includegraphics[clip, trim= 0cm 0.0cm 0.0cm 0.0cm,width=0.96\columnwidth]{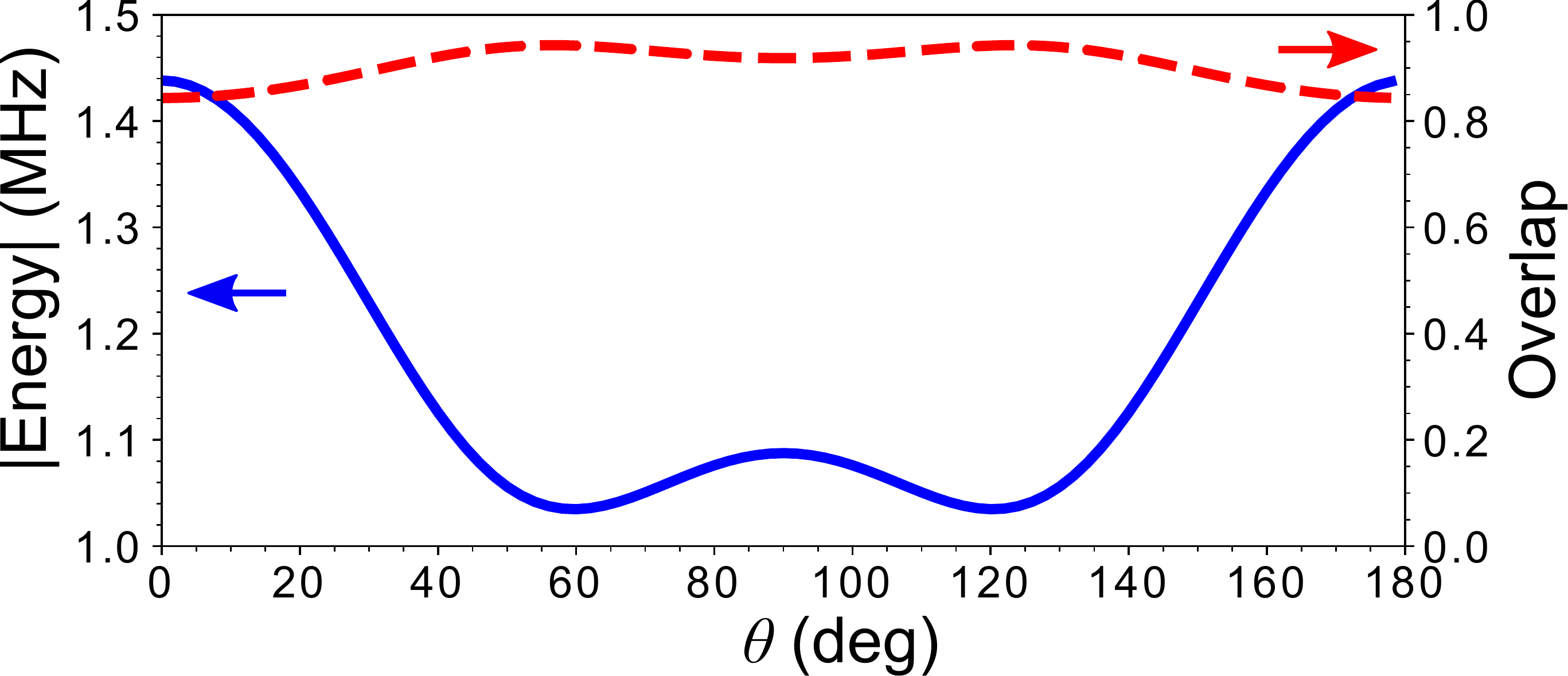}
\caption{\label{figZeeman3}%
Angular dependence of the adiabatic pair potential of $\ket{\text{Rb}92s_{1/2}\uparrow\text{K}95s_{1/2}\uparrow}$ caused by dipole-dipole interaction for $B$ = \SI{-5.5}{G} and $E$ = \SI{20}{mV/cm} at an atomic separation of \SI{22}{\mu m} (blue). The red curve shows the overlap with the eigenstate state $\ket{\text{Rb}92s_{1/2}\uparrow\text{K}95s_{1/2}\uparrow}$ for $V_\text{dd} = 0$, which is $>80\,\%$ for all angles $\theta$.  
}
\end{figure}%
\noindent
In Fig.~\ref{figZeeman2} we show pair potentials in the vicinity of $\ket{\alpha_a\alpha_b} = \ket{\text{Rb}92s_{1/2}\uparrow}\otimes\ket{\text{K}95s_{1/2}\uparrow}$ for different field configurations. Figures~\ref{figZeeman2}(a) and ~\ref{figZeeman2}(b) illustrate the case where no external $E$ and $B$ fields are present for two different angles (\SI{0}{\degree} and \SI{15}{\degree}). The states $\ket{\text{Rb}92s_{1/2}\text{K}95s_{1/2}}$ and $\ket{\text{Rb}92p_{1/2}\text{K}94p_{1/2}}$ are Zeeman degenerate for large $R$. Some of the Zeeman states, $\ket{\phi_i}$, show a weak interaction with $\ket{\alpha_a\alpha_b}$, where the overlap is encoded in the color map of the plot. For $B$ = 0 an extremely weak external electric field of less than \SI{10}{mV/cm} is necessary to obtain perfect resonance for the coupling of $\ket{\text{Rb}92s_{1/2}\text{K}95s_{1/2}}$ and $\ket{\text{Rb}92p_{1/2}\text{K}94p_{1/2}}$. However, the Zeeman degeneracy complicates the addressing of a single spin state which is needed to avoid interaction-limiting F\"orster zero states.
In the presence of both magnetic and electric fields, a scenario with multiple F\"orster resonances appears which can result in a flattening of the pair potential caused by compensating contributions from multiple states \cite{Paris-Mandoki2016}. As can be seen from Eq.~(\ref{eqrad}), channels with different $\Delta M$ have different angular dependencies and at nonzero angles all of these channels with $\Delta M $ = $0,\pm1,\pm2$ become dipole allowed. It is therefore important to choose electric and magnetic fields with care. In Figs.~\ref{figZeeman2}(c)--\ref{figZeeman2}(h) an external static magnetic field of $B$ = \SI{-5.5}{G} lifts the Zeeman degeneracy. For an additional electric field of \SI{20}{mV/cm} in Figs.~\ref{figZeeman2}(e) and \ref{figZeeman2}(f), the states are very well-isolated in energy and beyond \SI{10}{\mu m} all of the states that cross zero have a vanishing overlap with $\ket{\alpha_a\alpha_b}$. We find that for this combination of electric and magnetic fields a near-isotropic potential is realized, as depicted in Fig.~\ref{figZeeman3} for a interatomic distance of $R$ = \SI{22}{\mu m}. Energy shifts $>$ \SI{1.0}{\MHz} can be realized for all $\theta$ up to a distance of $\approx$ \SI{22}{\mu m}. Moreover, the pair state shows little mixing with other states and for all angles $|\braket{s\uparrow,s\downarrow| \phi_i}|^2 > 0.8$ is satisfied. Numerically calculated pair potentials for several angles ${\SI{0}{\degree} \leq \theta \leq \SI{180}{\degree}}$ are presented in Fig.~\ref{figAppendix} in Appendix~\ref{B} .  
A small change of the electric field magnitude to \SI{30}{\mV/cm}, however, can change the situation drastically, as shown in Figs.~\ref{figZeeman2}(g) and \ref{figZeeman2}(h). For $\theta = $ \SI{15}{\degree}, there is a significant overlap between $\ket{\alpha_a\alpha_b}$ and another pair states which is close in energy. As a consequence an excitation laser with a linewidth of \SI{1.2}{\MHz} could couple to more than a single Zeeman state at specific distances, \new{and compromising the blockade effect}, as described in Ref. \cite{Browaeys_2018b}. Consequently, the case presented in  Figs.~\ref{figZeeman2}(e) and \ref{figZeeman2}(f) is favorable for experimental applications, as we will discuss in the subsequent section. 
\newline
\section{APPLICATION: STRONG PHOTON-PHOTON INTERACTIONS} 
\label{sec5}
\begin{figure}[t] 
	\centering
	\includegraphics[clip, trim= 0.0cm 0.1cm 0.0cm 0.0cm,width=0.94\columnwidth]{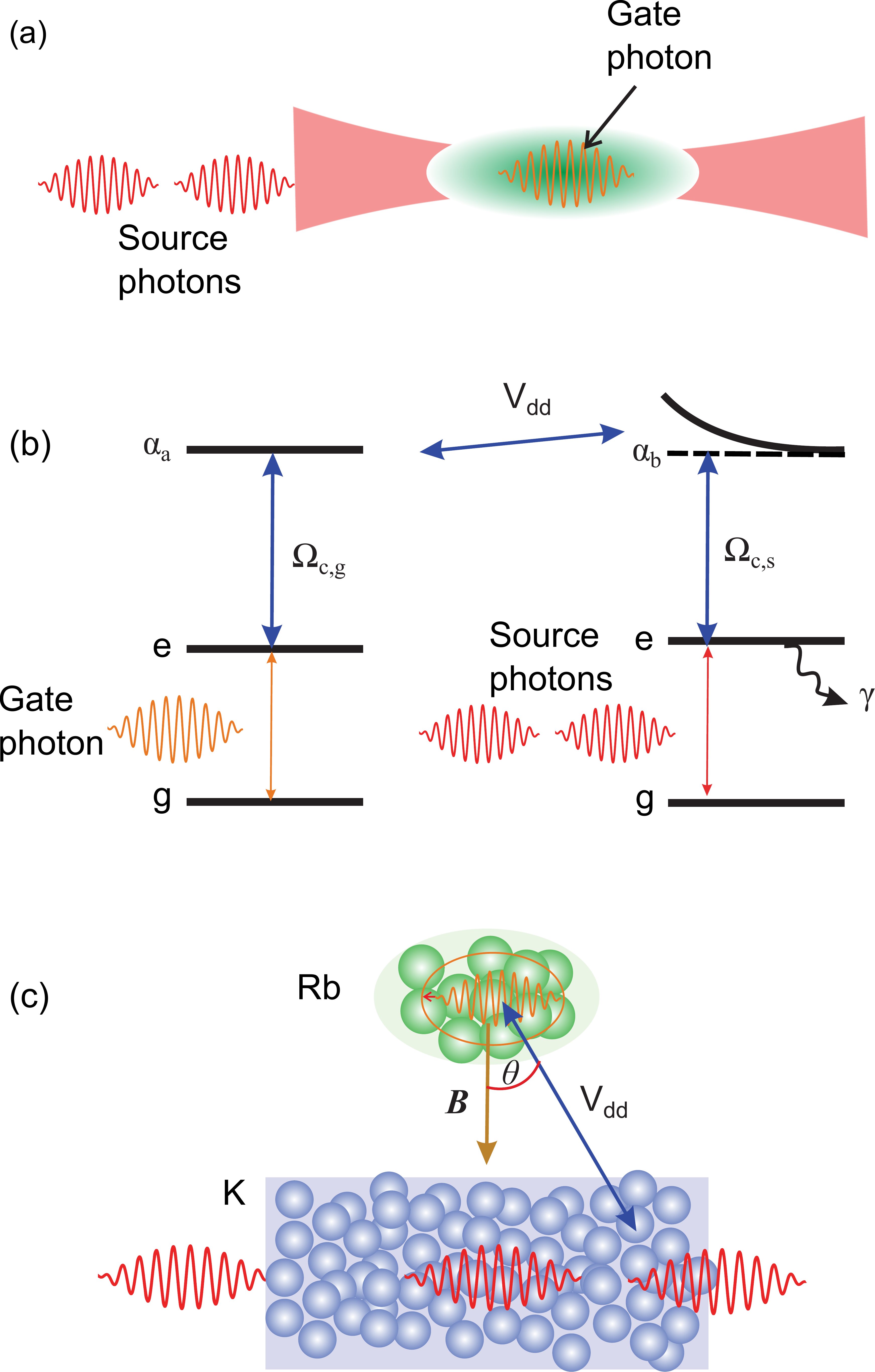}
	\caption{\label{figApplication}%
		(a) Operation of a Rydberg photonic transistor, where a single gate photon stored in an atomic ensemble controls the state of transmission of many source photons through the ensemble. (b) Optical excitation schemes used in a Rydberg single-photon transistor. A coupling laser beam with a Rabi frequency $\Omega_{c,g}$ is ramped down or up to store or retrieve a gate photon, while a second coupling laser beam with a Rabi frequency $\Omega_{c,s}$ establishes an EIT resonance condition for the source photons, which is altered by the presence of strong Rydberg-Rydberg interactions. (c) Spatially separated arrangement of Rb and K atomic ensembles where the former hosts the gate photon and the latter hosts the source photons. Here $\textbf{\textit{B}}$ shows the direction of the applied magnetic field which is taken as the quantization axis and $\theta$ is the angle between the interatomic axis of a specific pair of atoms and $\textit{\textbf{B}}$.
	}
\end{figure}%
The small zero-field F\"orster defects encountered above in the interspecies K-Rb Rydberg system present distinct advantages in applications that rely on strong, long-range Rydberg interactions and as a specific example, we consider its use in the context of single-photon optical transistors \cite{Tiarks2014,Gorniaczyk2014,Gorniaczyk2016,Firstenberg2016} where atoms mediate strong interactions between a single gate photon and a stream of source photons.

Figures~\ref{figApplication}(a) and \ref{figApplication}(b) illustrate the commonly realized operation of a Rydberg photon transistor which is based on a single trapped ensemble of a single atomic species. A gate photon is stored in the ensemble using an EIT scheme connecting a ground atomic state to a Rydberg state $|\alpha_a\rangle$, a so-called Rydberg polariton \cite{Firstenberg2016}. A stream of source photons then propagates through the medium under the conditions of EIT resonance connecting to a different Rydberg state $|\alpha_b\rangle$. In the absence of the gate photon, transmission of the source photons is maximum at the EIT resonance. When a gate photon is stored, however, the EIT resonance shifts due to the dipole-dipole interactions between $|\alpha_a\rangle$ and $|\alpha_b\rangle$ [Fig.~\ref{figApplication}(b)]. This causes the source photons to be, in principle, completely absorbed when the EIT resonance condition is met. The shift of the EIT resonance is significant only if the energy shift $\mathcal{U}_{\mathrm{dd}}$ due to dipole-dipole interactions exceeds the width of the EIT transmission window $\mathcal{U}_{\mathrm{EIT}} =  \Omega_{\mathrm{c}}^2/\gamma$, where $\Omega_{\mathrm{c}}$ is the Rabi frequency of the coupling transition in the source EIT ladder scheme and $\gamma$ is the natural linewidth of intermediate state. The distance over which this condition is fulfilled is called the photon blockade radius and the corresponding volume the blockade volume. 

Rydberg single-photon transistors have so far been implemented only using a single atomic species. Initial experimental demonstrations were achieved in Refs. \cite{Tiarks2014,Gorniaczyk2014} using ultracold Rb atoms, with a F\"orster resonance employed in Ref. \cite{Tiarks2014}. Stark-tuned F\"orster resonances in Rb were later used in Ref. \cite{Gorniaczyk2016} for enhanced contrast and gate photon coherence. Using F\"orster resonances for improving the performance of single-photon transistors, however, poses several challenges for single-species systems. First, the excitation hopping processes shown in Eq.~(\ref{hopping}) are present and near F\"orster resonances, are to the detriment for the operation of single-photon transistors \cite{Tiarks2014}. To suppress hopping, F\"orster resonances must be carefully chosen with small dipolar coupling strength for the alternative path, as has been done in Refs. \cite{Gorniaczyk2016, Paris-Mandoki2016}. This limits the number of useful F\"orster resonances and requires a compromise on the achievable photon blockade radius. Second, large transistor input photon rates require a small interaction between source photons, which is difficult to realize in single-species single-ensemble systems. For example, using the F\"orster resonance $|\alpha_a\alpha_b\rangle \leftrightarrow |\beta_a\beta_b\rangle = \ket{\text{Rb}84s_{1/2}\text{Rb}81s_{1/2}} \leftrightarrow \ket{\text{Rb}83p_{1/2} \text{Rb}81p_{1/2}}$ shown in Table~\ref{table_comp} and assuming $\mathcal{U}_{\mathrm{EIT}}/2\pi = 1.1$\,MHz, one obtains a photon blockade radius of 11\,$\mu$m for the pair state $\ket{\text{Rb}84s_{1/2},m_j = 1/2}\otimes\ket{\text{Rb}81s_{1/2},m_j = 1/2}$ at ${\theta = 0\degree}$ at zero-field. In comparison, the photon blockade radius for the $\ket{\text{Rb}84s_{1/2},m_j = 1/2}\otimes\ket{\text{Rb}84s_{1/2},m_j = 1/2}$ pair state is about 8\,$\mu$m, so the unwanted interaction between source photons is significant within this distance. The relatively small photon blockade volume for the gate-source interaction means that achieving large optical depth within the blockade volume ($\mathrm{OD}_{\mathrm{}b}$) for the source photons can only be achieved with high atomic densities. Rydberg-ground state collisions at dense atomic gases, however, cause severe decoherence and losses and has been identified as the most critical bottleneck for coherent operation of single-photon transistors \cite{Firstenberg2016,Gorniaczyk2016}. Furthermore, the F\"orster defect for the aforementioned resonance is $2\pi\times 3$\,MHz and a relatively large electric field of $\sim 1$\,V/cm is required to bring the pair states to exact resonance. For high $n$ Rydberg states, such fields can cause many dipole-allowed molecular states to come into play at short ($\sim 1\,\mu$m) distances and leads to a breakdown of blockade.

Interspecies, strong, near-zero field F\"orster resonances discussed in this paper can offer solutions to these problems. First, as noted in Sec.~\ref{sec3C}, the excitation hopping process, Eq.~(\ref{hopping}), is absent for interspecies F\"orster resonances, and a wide array of such resonances become suitable choices. Considering the specific F\"orster resonance $|\alpha_a\rangle \otimes |\alpha_b\rangle \leftrightarrow |\beta_a\rangle \otimes |\beta_b\rangle = \ket{\text{Rb}92s_{1/2}}\otimes \ket{\text{K}95s_{1/2}} \leftrightarrow \ket{\text{Rb}92p_{1/2}}\otimes\ket{\text{K}94p_{1/2}}$ discussed in detail in Sec.~\ref{sec4}, we note that even at zero electric field the photon blockade volume is very large. In particular, assuming the initial state to be $\ket{\text{Rb}92s_{1/2},m_j = 1/2}\otimes\ket{\text{K}95s_{1/2},m_j = 1/2}$ and a typical EIT linewidth $\mathcal{U}_{\mathrm{EIT}}/2\pi = 1.1\,$MHz, one obtains a photon blockade radius of $r_b \geq 22 \,\mu$m over the entire angular range with a small applied field of magnitude 20\,mV/cm (Fig.~\ref{figZeeman3}). The corresponding blockade volume is larger than the size of atomic ensembles typically realized in ultracold atomic experiments, allowing for a large atomic ensemble to fit entirely within the blockade volume to host the source photons. As pointed out in Ref. \cite{Li2015}, the scattering of source photons within this volume can not provide information about the position of the gate photon, and as a result the coherence of the gate photon is preserved in the scattering process. 

The large photon blockade radius allows one to envisage the arrangement shown in Fig.~\ref{figApplication}(c) for realizing a single-photon transistor, where atomic ensembles hosting the gate photon and the source photons are spatially isolated. Specifically, a 780\,nm gate photon is stored in a dilute Rb ensemble, whereas the 767\,nm source photons propagate through a separate cylinder-shaped ensemble of K atoms with a radius $\sigma_r = 6\,\mu$m and length $\sigma_z = 40\,\mu$m, and atom number $N = 5000$, placed at a center-to-center distance of $17\,\mu$m from the gate ensemble. A fully blockaded ensemble with an optical density $\mathrm{OD}_\mathrm{b} = \rho\zeta\sigma_z \sim 22$ is realized, where $\lambda = $ 767\,nm and $\zeta = 3\lambda^2/2\pi$ is the on-resonance photon scattering crosssection. This, in principle, realizes a near-perfect photon switching with a contrast of $[1-\exp(-\mathrm{OD}_\mathrm{b})]$ depending on whether a gate photon is absent or present at zero electric field, while keeping the coherence of the gate photon preserved against source photon scattering. Moreover, the interaction between source photons is sufficiently small ($\lesssim 2\pi\times 200$ kHz at 20\,$\mu$m), ensuring the correlation between two source photons remains small at such distances. Because of the spatial separation between the two ensembles in our proposed arrangement, short-distance molecular potential curves between K and Rb are completely avoided.

\section{CONCLUSION} 
\label{sec6}
We have investigated the dipole-dipole interaction between rubidium and potassium atoms in their Rydberg states for a wide range of principal quantum numbers, and in particular we considered pair states comprising angular momentum channels that are amenable to two-photon Rydberg excitations. Using large spin-independent interaction coefficients $C_{3k}$ and small F\"orster defects as the figures of merit, we identified several strong F\"orster resonances with extremely small F\"orster defects at zero field. This results in strong ($>$\SI{1}{\MHz}) Rydberg-Rydberg interactions for distances exceeding $\SI{25}{\mu m}$ and very large crossover distances ($\approx\SI{100}{\mu m}$). When comparing F\"orster resonances in K-Rb with other frequently used atomic species combinations, Rb-Rb, Rb-Cs, and Cs-Cs, we found that K-Rb has a unique range of strong F\"orster resonances with near-zero F\"orster defects and large $C_{3k}$. Investigations of pair potentials of the F\"orster resonant pair states suggest that by judiciously choosing Zeeman states for the initial pair states and small magnitudes of the applied fields, one can obtain strong blockade interactions over the entire angular range.


Interspecies zero field F\"orster resonances in the K-Rb system opens up new opportunities and directions for photonic devices that rely on strong Rydberg atom-mediated photon-photon interactions, such as single-photon Rydberg transistors and photonic quantum gates \cite{Tiarks2018}. Our analysis suggests that multiple, very large ensembles that are spatially separated can be considered for such applications, which would eliminate the strong density-dependent collisional losses while enabling large optical depth (OD) within the blockade volume. As pointed out in Ref. \cite{Firstenberg2016}, realizing large $\mathrm{OD}_{\mathrm{}b}$ without incurring losses and decoherence due to inelastic collisions between Rydberg electrons and ground state atoms in dense ensembles \cite{Schlagmuller_2016} is a significant impediment in Rydberg quantum optics. \new{The K-Rb F\"orster resonances discussed here offer a solution to this by allowing very large ensembles that are spatially separated to be used.}

Beyond Rydberg photonic devices, strong interspecies F\"orster resonances have been proposed as a suitable tool for quantum non-demolition detection of qubit states for Rydberg atom-based quantum information processing \cite{Beterov2015}. More recently, interspecies Rydberg-Rydberg interactions with two isotopes of Rb have been employed to realize quantum gates with minimal cross-talk \cite{Zeng2017}. Our findings suggest that the K-Rb F\"orster resonances discussed in this work will be readily suitable for such applications. The relatively slow $1/R^3$ variation of energy shifts over distances $\sim 100\,\mu$m near K-Rb F\"orster resonances considered here can also offer useful benefits for long-range coherent excitation transfer schemes that involve adiabatic passage across a Rydberg state \cite{Ravets2014,Petrosyan2018}. Finally, the use of fermionic $^{40}$K combined with the bosonic species $^{87}$Rb paves the way to explore phenomena related to quantum statistics and degeneracy of atoms \cite{Xiaopeng2015}.


\begin{acknowledgments}
We acknowledge funding from the Marsden Fund of New Zealand (Contract No. UOO1729) and MBIE (Contract No. UOOX1915). \new{We thank Ryan Thomas for valuable comments on our paper.}
\end{acknowledgments}


\appendix
\section{CALCULATION OF INTERACTION PARAMETERS}
\label{A}
\new{Our PYTHON code employs functions from the \textit{Alkali Rydberg Calculator} (ARC) toolbox from Durham University \cite{Durham_tutorial}. This allows us to calculate several features of Rydberg atoms using the framework of quantum defect theory (QDT) with the Numerov integration method in the Coulomb approximation for the radial part of the wave functions. We have modified the code, where needed, to allow for interspecies pair states. $C_{3k}$ and $C_{6k}$ interaction coefficients are calculated using radial wave functions, dipole matrix elements, and reduced matrix elements for a K-Rb interspecies system, as well as for variable combinations of all alkali atoms. For the evaluation of external fields with arbitrary directions relative to the quantization axis, we make use of the open-source \textit{Pair Interaction} tool from Stuttgart \cite{Weber2017}. This tool is used when several pair states have to be taken into account and the Hamiltonian has to be diagonalized numerically.}

\section{ADDITIONAL PAIR POTENTIALS FOR THE PAIR STATE $\ket{\text{Rb}92s_{1/2}\uparrow}\otimes \ket{\text{K}95s_{1/2}\uparrow}$ }
\label{B}
\begin{figure*}[t!] 
	\includegraphics[clip, trim= 0.0cm 0cm 0cm 0cm,width=\textwidth]{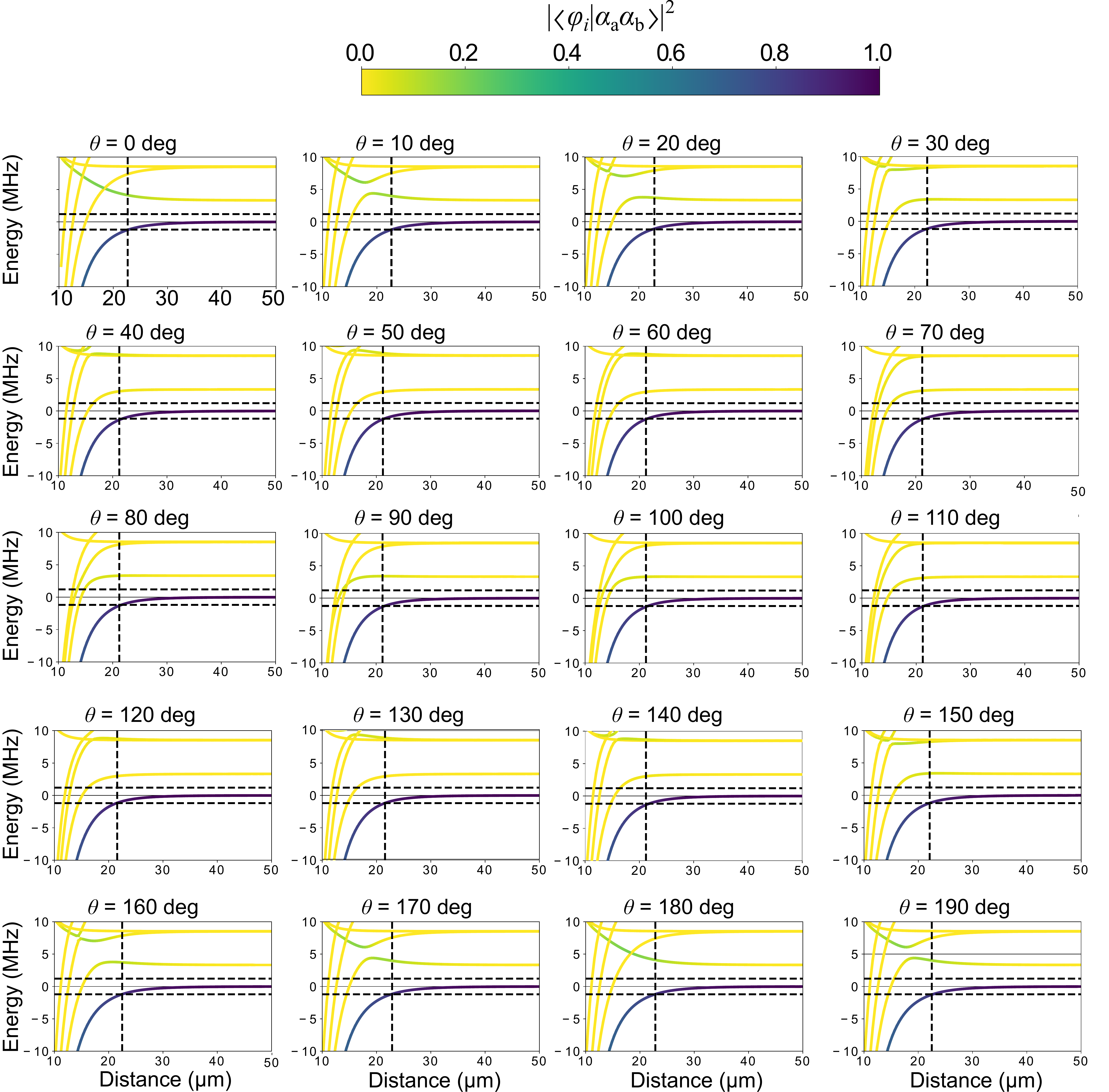}%
	\caption{Pair potentials in the vicinity of $\ket{\alpha_a \alpha_b} = \ket{\text{Rb}92s_{1/2}\uparrow}\otimes \ket{\text{K}95s_{1/2}\uparrow}$ for static fields of $B_z$ = \SI{-5.5}{G} and $E_z$ = \SI{20}{mV/cm} for angles $\theta$ between \SI{0}{\degree} and \SI{190}{\degree}. For all angles, the shift in energy is $>$ \SI{1.2}{\MHz} (horizontal dashed lines) for $R \lesssim \SI{22}{\mu m}$, resulting in a near-isotopic potential. The color code depicts the overlap with $\ket{\text{Rb}92s_{1/2}\uparrow}\otimes \ket{\text{K}95s_{1/2}\uparrow}$. Also see Fig.~\ref{figZeeman2} in Sec.~\ref{sec4}.}
	\label{figAppendix}
\end{figure*}

In Fig.~\ref{figAppendix} we show an extension of Fig.~\ref{figZeeman2}(e) and (f) for angles between $0\degree$ and $190\degree$. All pair potentials show an overlap $> 80\%$ with the eigenstate $\ket{\text{Rb}92s_{1/2}\uparrow} \otimes \ket{\text{K}95s_{1/2}\uparrow}$, and nearly no mixing with other states. In addition, energy shifts $>1.0\,$MHz can be obtained up to an interatomic distance of $\approx 22\,\mu m$ for all angles. These pair potentials confirm that for the particular combination of electric and magnetic fields, the interaction potential is near-isotropic.
\onecolumngrid
\clearpage
\section{TABLES WITH A SELECTION OF K-K and K-Rb PAIR STATE COMBINATIONS }
In this Appendix, we list pair state combinations with small F\"orster defects for the two-species K-Rb system in Table~\ref{table_s}, and for the single-species system K-K in Table~\ref{table_KK}. The pair combinations in Table~\ref{table_s} correspond to the dark points in Figs.~\ref{figS_state} and \ref{fig_d}. A complete set of all pair combinations and the corresponding state quantum numbers, F\"orster defects and interaction coefficients can be found in the Supplemental Material \cite{Supplemental}. 
\label{C}



\begin{table*}[htb!] 
\caption{ \label{table_s}A selection of K-Rb pair states with F\"orster defects $|\Delta|/(2\pi) <$ \SI{12}{\MHz} and $|C_{3k}| >$ \SI{1}{\GHz\: \um^3} for principal quantum numbers $50  \leq n \leq 110$. $R_c$ is calculated using the strongest coefficient $\mathcal{S}_k$ (Eq.~(\ref{ceq2})), e.g. see \cite{Beterov2015,Walker_Saffman}. \new{For a complete set of all pair combinations of Figs.~\ref{figS_state} and \ref{fig_d} and corresponding $C_{3k}$ and $\Delta$ see Supplemental Material at \cite{Supplemental}.} 
}
\begin{ruledtabular}
\begin{tabular}{ccccccc}

\textbf{ss $\rightarrow$ pp }\\
      Rb initial & Rb final & K initial & K final & $\Delta/2\pi$ (MHz) & $C_{3k}$ (\SI{}{\GHz\: \um^3})& $R_c$ (\SI{}{\um})  \\ 
 $52s_{1/2}$ &$52p_{3/2}$ &$53s_{1/2}$ &$52p_{1/2}$ & 1.85 & 2.37 &  12.2 \\     
 $56s_{1/2}$ &$56p_{3/2}$ &$57s_{1/2}$ &$56p_{3/2}$ & -11.71 & -3.25  & 8.1 \\ 
 $59s_{1/2}$ &$59p_{3/2}$ &$85s_{1/2}$ &$83p_{3/2}$ & -4.40 & 1.21  & 8.1 \\ 
 $60s_{1/2}$ &$59p_{3/2}$ &$57s_{1/2}$ &$57p_{3/2}$ & -7.50 & -3.95& 10.1 \\ 
 $61s_{1/2}$ &$61p_{3/2}$ &$88s_{1/2}$ &$86p_{1/2}$ & 6.46 & -1.4 &  6.7 \\ 
 $66s_{1/2}$ &$66p_{1/2}$ &$68s_{1/2}$ &$67p_{1/2}$ & 2.50 & -6.69 &  15.3 \\ 
 $69s_{1/2}$ &$69p_{1/2}$ &$71s_{1/2}$ &$70p_{3/2}$ & 9.88 & 8.01 &  10.5 \\ 
 $70s_{1/2}$ &$70p_{1/2}$ &$72s_{1/2}$ &$71p_{3/2}$ & -7.08 & 8.56 & 12.0 \\ 
 $71s_{1/2}$ &$70p_{1/2}$ &$67s_{1/2}$ &$67p_{1/2}$ & -7.41 & -7.76 &  11.2 \\ 
 $72s_{1/2}$ &$72p_{1/2}$ &$105s_{1/2}$ &$103p_{1/2}$ & -0.20 & 2.89 &  26.8 \\ 
 $74s_{1/2}$ &$73p_{1/2}$ &$70s_{1/2}$ &$70p_{3/2}$ & 7.67 & 9.23 &  11.9 \\     
 $75s_{1/2}$ &$74p_{1/2}$ &$71s_{1/2}$ &$71p_{3/2}$ & -7.37 & 9.78 & 12.3 \\ 
 $85s_{1/2}$ &$85p_{3/2}$ &$87s_{1/2}$ &$86p_{1/2}$ & 7.82 & 18.56 &  15.0 \\ 
 $86s_{1/2}$ &$86p_{3/2}$ &$88s_{1/2}$ &$87p_{1/2}$ & 1.71 & 19.47  & 25.2 \\ 
 $87s_{1/2}$ &$87p_{3/2}$ &$89s_{1/2}$ &$88p_{1/2}$ & -3.85 & 20.41  & 19.6 \\ 
 $88s_{1/2}$ &$87p_{3/2}$ &$84s_{1/2}$ &$84p_{1/2}$ & 8.78& 19.70 & 14.7 \\ 
 $88s_{1/2}$ &$88p_{3/2}$ &$90s_{1/2}$ &$89p_{3/2}$ & -8.90 & 21.38 & 15.0 \\ 
 $90s_{1/2}$ &$89p_{3/2}$ &$86s_{1/2}$ &$86p_{1/2}$ & -2.97 & 21.66  & 21.8 \\ 
 $92s_{1/2}$ &$92p_{1/2}$ &$95s_{1/2}$ &$94p_{1/2}$ & 0.01 & -26.61 & 166.4 \\ 
 $93s_{1/2}$ &$93p_{3/2}$ &$95s_{1/2}$ &$94p_{3/2}$ & -2.50 & -26.90& 27.5 \\ 
 $96s_{1/2}$ &$95p_{3/2}$ &$92s_{1/2}$ &$92p_{3/2}$ & -0.11 & -28.32&  78.4 \\ 
 $97s_{1/2}$ &$96p_{1/2}$ &$92s_{1/2}$ &$92p_{1/2}$ & -0.10 & -28.53&  72.6 \\ 
 $97s_{1/2}$ &$97p_{1/2}$ &$100s_{1/2}$ &$99p_{3/2}$ & -0.63 & 33.12&   42.1 \\ 
 $102s_{1/2}$ &$101p_{1/2}$ &$97s_{1/2}$ &$97p_{3/2}$ & 1.59 & 31.15 & 31.5\\ 
 $103s_{1/2}$ &$102p_{1/2}$ &$98s_{1/2}$ &$98p_{3/2}$ & -2.38 & 36.62& 27.9 \\ 
            \\
            \\

\textbf{dd $\rightarrow$ pp, pf, ff}\\
      Rb initial & Rb final & K initial & K final & $\Delta/2\pi$ (MHz) & $C_{3k}$ (\SI{}{\GHz\: \um^3})&   $R_c$ (\SI{}{\um})  \\ 
 $53d_{3/2}$ &$54p_{1/2}$ &$63d_{3/2}$ &$65p_{1/2}$ & -9.38& -9.24&  9.9 \\ 
 $59d_{3/2}$ &$60p_{1/2}$ &$55d_{3/2}$ &$55f_{5/2}$ & -3.65& 9.46& 14.3 \\       
 $59d_{3/2}$ &$58f_{5/2}$ &$76d_{3/2}$ &$75f_{5/2}$ & 6.94 & -7.13& 12.1 \\           
 $60d_{3/2}$ &$59f_{5/2}$ &$65d_{3/2}$ &$66p_{1/2}$ & -1.85& 11.66&   19.3 \\ 
 $61d_{3/2}$ &$59f_{5/2}$ &$56d_{3/2}$ &$58p_{1/2}$ & -0.91& 4.53&  18.8 \\   
 $62d_{5/2}$ &$60f_{5/2}$ &$57d_{5/2}$ &$59p_{3/2}$ & 6.30 & -1.22&   9.9 \\ 
 $71d_{5/2}$ &$70f_{5/2}$ &$77d_{5/2}$ &$78p_{3/2}$ & 1.26 & -5.88 & 28.4 \\ 
 $72d_{3/2}$ &$74p_{1/2}$ &$73d_{3/2}$ &$72f_{5/2}$ & 4.56& 7.59& 12.4 \\                    
 $81d_{3/2}$ &$82p_{1/2}$ &$76d_{3/2}$ &$76f_{5/2}$ & 0.82& 34.41&  36.3 \\     
 $82d_{5/2}$ &$83p_{3/2}$ &$100d_{5/2}$ &$102p_{3/2}$ & 0.29& -51.54&   59.0 \\
 $83d_{3/2}$ &$81f_{5/2}$ &$60d_{3/2}$ &$60f_{5/2}$ & -0.35& -10.66& 37.3 \\    
 $91d_{3/2}$ &$90f_{5/2}$ &$99d_{3/2}$ &$100p_{1/2}$ &1.71& 62.91&  34.7 \\          
 $92d_{5/2}$ &$91f_{5/2}$ &$100d_{5/2}$ &$101p_{3/2}$ & 0.54& -16.72&  53.3 \\     
 $99d_{5/2}$ &$97f_{5/2}$ &$92d_{5/2}$ &$94p_{3/2}$ & -0.79& -8.46& 37.5 \\     
 $102d_{3/2}$ &$100f_{5/2}$ &$74d_{3/2}$ &$74f_{5/2}$ & -1.06& -24.78&   34.1\\         
 $103d_{5/2}$ &$105p_{3/2}$ &$104d_{5/2}$ &$103f_{5/2}$ & 0.28& -8.05& 52.4 \\  
 $103d_{3/2}$ &$104p_{1/2}$ &$97d_{3/2}$ &$97f_{5/2}$ & 0.72& 91.16 & 52.4 \\ 
 $104d_{3/2}$ &$105p_{1/2}$ &$98d_{3/2}$ &$98f_{5/2}$ & -1.90& 94.88& 38.5 \\   
 \end{tabular}
\end{ruledtabular}
\end{table*}
\begin{table*}
\caption{Continued from Table~\ref{table_s}.}
\begin{ruledtabular}
\begin{tabular}{ccccccc}

\textbf{sd $\rightarrow$ pp, pf}\\
      Rb initial &  K initial &Rb final & K final & $\Delta/2\pi$ (MHz) & $C_{3k}$ (\SI{}{\GHz\: \um^3})&   $R_c$ (\SI{}{\um})  \\ 
57$s_{1/2}$  & 55$d_{5/2}$  & 56$p_{1/2}$  & 57$p_{3/2}$  & 2.06   & 3.99    & 13.71 \\
60$s_{1/2}$  & 56$d_{3/2}$  & 60$p_{1/2}$  & 57$p_{1/2}$  & 1.28   & 6.4        & 17.94 \\
60$s_{1/2}$  & 66$d_{3/2}$  & 60$p_{3/2}$  & 65$f_{5/2}$  & 2.6    & 4.02       & 14.56 \\
60$s_{1/2}$  & 81$d_{3/2}$  & 59$p_{1/2}$  & 84$p_{1/2}$  & 3.05   & 1.7        & 8.63  \\
61$s_{1/2}$  & 59$d_{3/2}$  & 60$p_{1/2}$  & 61$p_{1/2}$  & 2.69   & 5.62       & 13.42 \\
67$s_{1/2}$  & 74$d_{5/2}$  & 67$p_{3/2}$  & 73$f_{5/2}$  & 0.73   & 1.71      & 15.83 \\
69$s_{1/2}$  & 64$d_{5/2}$  & 69$p_{3/2}$  & 65$p_{3/2}$  & 4.49   & 10.49      & 15.62 \\
69$s_{1/2}$  & 77$d_{3/2}$  & 69$p_{1/2}$  & 76$f_{5/2}$  & 9.34   & 7.51       & 10.95 \\
70$s_{1/2}$  & 65$d_{3/2}$  & 70$p_{3/2}$  & 66$p_{3/2}$  & 10.93  & 3.72       & 8.08  \\
72$s_{1/2}$  & 71$d_{3/2}$  & 71$p_{3/2}$  & 73$p_{3/2}$  & 1.08   & 3.72       & 17.5  \\
73$s_{1/2}$  & 68$d_{3/2}$  & 73$p_{3/2}$  & 69$p_{1/2}$  & 7.29   & 13.99     & 13.95 \\
74$s_{1/2}$  & 57$d_{3/2}$  & 73$p_{3/2}$  & 57$f_{5/2}$  & 2.94   & 8.67      & 18.06 \\
74$s_{1/2}$  & 69$d_{3/2}$  & 74$p_{3/2}$  & 70$p_{1/2}$  & 7.78   & 14.82     & 13.91 \\
74$s_{1/2}$  & 82$d_{5/2}$  & 74$p_{3/2}$  & 81$f_{5/2}$  & 1.99   & 2.6        & 13.03 \\
75$s_{1/2}$  & 74$d_{3/2}$  & 74$p_{3/2}$  & 76$p_{1/2}$  & 8.69   & 13.99     & 13.15 \\
76$s_{1/2}$  & 75$d_{3/2}$  & 75$p_{3/2}$  & 77$p_{1/2}$  & 3.48   & 14.78     & 18.17 \\
77$s_{1/2}$  & 108$d_{5/2}$ & 77$p_{1/2}$  & 108$p_{3/2}$ & 0.15   & 3.55      & 31.81 \\
81$s_{1/2}$  & 90$d_{5/2}$  & 81$p_{3/2}$  & 89$f_{5/2}$  & 3.09   & 3.79       & 12.75 \\
86$s_{1/2}$  & 51$d_{3/2}$  & 89$p_{1/2}$  & 50$f_{5/2}$  & 0.02   & 0.13        & 22.91 \\
87$s_{1/2}$  & 82$d_{5/2}$  & 87$p_{1/2}$  & 83$p_{3/2}$  & 5.2    & 28.32     & 19.37 \\
87$s_{1/2}$  & 82$d_{5/2}$  & 87$p_{1/2}$  & 83$p_{3/2}$  & 5.2    & 28.32     & 19.37 \\
88$s_{1/2}$  & 83$d_{5/2}$  & 88$p_{1/2}$  & 84$p_{3/2}$  & 0.76   & 29.71   & 37.32 \\
88$s_{1/2}$  & 83$d_{5/2}$  & 88$p_{1/2}$  & 84$p_{3/2}$  & 0.76   & 29.71    & 37.32 \\
89$s_{1/2}$  & 84$d_{3/2}$  & 89$p_{1/2}$  & 85$p_{3/2}$  & 4.07   & 10.38     & 14.78 \\
90$s_{1/2}$  & 85$d_{3/2}$  & 90$p_{1/2}$  & 86$p_{3/2}$  & 9.06   & 10.88      & 11.5  \\
93$s_{1/2}$  & 92$d_{5/2}$  & 92$p_{1/2}$  & 94$p_{3/2}$  & 6.45   & 31.52     & 18.68 \\
94$s_{1/2}$  & 93$d_{5/2}$  & 93$p_{1/2}$  & 95$p_{3/2}$  & 2.21   & 32.93     & 27.08 \\
94$s_{1/2}$  & 88$d_{5/2}$  & 94$p_{3/2}$  & 89$p_{3/2}$  & 2.23   & 37.66    & 30.19 \\
94$s_{1/2}$  & 89$d_{3/2}$  & 94$p_{1/2}$  & 90$p_{1/2}$  & 4.31   & 41.12     & 22.25 \\
94$s_{1/2}$  & 105$d_{3/2}$ & 94$p_{3/2}$  & 104$f_{5/2}$ & 12.98  & 26.38    & 15.96 \\
95$s_{1/2}$  & 94$d_{3/2}$  & 94$p_{1/2}$  & 96$p_{3/2}$  & 0.17   & 11.47    & 43.86 \\
95$s_{1/2}$  & 90$d_{3/2}$  & 95$p_{1/2}$  & 91$p_{1/2}$  & 0.29   & 42.97    & 55.36 \\
95$s_{1/2}$  & 107$d_{3/2}$ & 95$p_{1/2}$  & 106$f_{5/2}$ & 0.47   & 28.47    & 46.26 \\
95$s_{1/2}$  & 89$d_{3/2}$  & 95$p_{3/2}$  & 90$p_{3/2}$  & 1.64   & 13.12    & 23.16 \\
96$s_{1/2}$  & 91$d_{3/2}$  & 96$p_{1/2}$  & 92$p_{1/2}$  & 3.38   & 44.89    & 24.83 \\
96$s_{1/2}$  & 96$d_{5/2}$  & 95$p_{3/2}$  & 98$p_{3/2}$  & 18.72  & 37.34    & 14.82 \\
97$s_{1/2}$  & 92$d_{3/2}$  & 97$p_{1/2}$  & 93$p_{1/2}$  & 6.75   & 46.88    & 20.02 \\
98$s_{1/2}$  & 98$d_{5/2}$  & 97$p_{3/2}$  & 100$p_{3/2}$ & 8.66   & 40.63    & 19.72 \\
98$s_{1/2}$  & 93$d_{3/2}$  & 98$p_{1/2}$  & 94$p_{1/2}$  & 9.83   & 48.92     & 17.91 \\
98$s_{1/2}$  & 92$d_{3/2}$  & 98$p_{3/2}$  & 93$p_{1/2}$  & 10.35  & 47.12     & 18.6  \\
99$s_{1/2}$  & 99$d_{5/2}$  & 98$p_{3/2}$  & 101$p_{3/2}$ & 4.2    & 42.35    & 25.44 \\
99$s_{1/2}$  & 93$d_{3/2}$  & 99$p_{3/2}$  & 94$p_{1/2}$  & 5.53   & 49.16    & 23.25 \\
99$s_{1/2}$  & 98$d_{3/2}$  & 98$p_{1/2}$  & 100$p_{1/2}$ & 8.19   & 43.17    & 18.26 \\
100$s_{1/2}$ & 100$d_{5/2}$ & 99$p_{3/2}$  & 102$p_{3/2}$ & 0.09   & 44.12   & 92.48 \\
100$s_{1/2}$ & 94$d_{3/2}$  & 100$p_{3/2}$ & 95$p_{1/2}$  & 1.09   & 51.28    & 40.48 \\
100$s_{1/2}$ & 99$d_{3/2}$  & 99$p_{1/2}$  & 101$p_{1/2}$ & 5.02   & 44.99    & 21.79 \\
101$s_{1/2}$ & 100$d_{3/2}$ & 100$p_{1/2}$ & 102$p_{1/2}$ & 2.09   & 46.86    & 29.56 \\
101$s_{1/2}$ & 101$d_{3/2}$ & 100$p_{3/2}$ & 103$p_{3/2}$ & 2.49   & 15.32    & 21.21 \\
101$s_{1/2}$ & 95$d_{3/2}$  & 101$p_{3/2}$ & 96$p_{1/2}$  & 2.99   & 53.46    & 29.36 \\
102$s_{1/2}$ & 101$d_{3/2}$ & 101$p_{1/2}$ & 103$p_{1/2}$ & 0.61   & 48.79   & 45.27 \\
102$s_{1/2}$ & 96$d_{3/2}$  & 102$p_{3/2}$ & 97$p_{1/2}$  & 6.74   & 55.71   & 22.7  \\
103$s_{1/2}$ & 102$d_{3/2}$ & 102$p_{1/2}$ & 104$p_{1/2}$ & 3.1    & 50.79   & 26.64 \\
103$s_{1/2}$ & 103$d_{3/2}$ & 102$p_{3/2}$ & 105$p_{1/2}$ & 8.91   & 52.73   & 20.3  \\
104$s_{1/2}$ & 104$d_{3/2}$ & 103$p_{3/2}$ & 106$p_{1/2}$ & 5.4    & 54.85    & 24.3  \\
104$s_{1/2}$ & 103$d_{3/2}$ & 103$p_{1/2}$ & 105$p_{1/2}$ & 5.4    & 52.84    & 22.44 \\
105$s_{1/2}$ & 82$d_{5/2}$  & 104$p_{3/2}$ & 82$f_{5/2}$  & 0.54   & 9.94     & 31.48 \\
105$s_{1/2}$ & 105$d_{3/2}$ & 104$p_{3/2}$ & 107$p_{1/2}$ & 2.15   & 57.04    & 33.46 \\
106$s_{1/2}$ & 106$d_{3/2}$ & 105$p_{3/2}$ & 108$p_{1/2}$ & 0.86   & 59.29    & 46.08 \\
107$s_{1/2}$ & 107$d_{3/2}$ & 106$p_{3/2}$ & 109$p_{1/2}$ & 3.64   & 61.6     & 28.81 \\
108$s_{1/2}$ & 107$d_{3/2}$ & 107$p_{1/2}$ & 109$p_{1/2}$ & 12.92  & 61.68    & 17.67 \\
110$s_{1/2}$ & 86$d_{3/2}$  & 109$p_{3/2}$ & 86$f_{5/2}$  & 1.34   & 45.04   & 40.66 \\
\\
\end{tabular}
\end{ruledtabular}
\end{table*}
\begin{table*}

\caption{Continued from Table~\ref{table_s}}
\begin{ruledtabular}
\begin{tabular}{ccccccc}
\textbf{ds $\rightarrow$ pp, fp}\\
      Rb initial &  K initial & Rb final &K final & $\Delta/2\pi$ (MHz) & $C_{3k}$ (\SI{}{\GHz\: \um^3})&  $R_c$ (\SI{}{\um})  \\ 
52$d_{3/2}$  & 60$s_{1/2}$  & 53$p_{1/2}$  & 60$p_{1/2}$  & 11.793 & -6.21      & 8.47  \\
52$d_{5/2}$  & 62$s_{1/2}$  & 51$f_{7/2}$  & 61$p_{3/2}$  & 0.539  & -4.76     & 26.04 \\
57$d_{3/2}$  & 51$s_{1/2}$  & 55$f_{5/2}$  & 51$p_{1/2}$  & 3.503  & 2.49      & 10.51 \\
57$d_{5/2}$  & 68$s_{1/2}$  & 56$f_{7/2}$  & 67$p_{1/2}$  & 9.146  & 6.92      & 10.73 \\
58$d_{3/2}$  & 52$s_{1/2}$  & 56$f_{5/2}$  & 52$p_{3/2}$  & 5.757  & -2.68     & 9.77  \\
58$d_{3/2}$  & 68$s_{1/2}$  & 59$p_{3/2}$  & 68$p_{3/2}$  & 2.651  & 3.2       & 12.32 \\
61$d_{3/2}$  & 57$s_{1/2}$  & 63$p_{3/2}$  & 56$p_{3/2}$  & 0.406  & 1.26      & 16.89 \\
64$d_{3/2}$  & 75$s_{1/2}$  & 65$p_{3/2}$  & 75$p_{3/2}$  & 4.016  & 4.78      & 12.26 \\
64$d_{3/2}$  & 85$s_{1/2}$  & 66$p_{1/2}$  & 83$p_{3/2}$  & 4.605  & -1.51     & 7.73  \\
65$d_{3/2}$  & 61$s_{1/2}$  & 67$p_{1/2}$  & 60$p_{3/2}$  & 2.917  & 5.38      & 13.77 \\
65$d_{3/2}$  & 76$s_{1/2}$  & 66$p_{3/2}$  & 76$p_{1/2}$  & 6.925  & -5.09      & 9.76  \\
66$d_{5/2}$  & 85$s_{1/2}$  & 64$f_{7/2}$  & 86$p_{3/2}$  & 2.221  & 1.15       & 10.1  \\
67$d_{3/2}$  & 60$s_{1/2}$  & 65$f_{5/2}$  & 60$p_{3/2}$  & 3.088  & -4.87     & 14.67 \\
70$d_{3/2}$  & 82$s_{1/2}$  & 71$p_{3/2}$  & 82$p_{3/2}$  & 6.715  & 6.89      & 11.67 \\
71$d_{3/2}$  & 82$s_{1/2}$  & 72$p_{1/2}$  & 82$p_{3/2}$  & 2.037  & 22.21     & 24.89 \\
71$d_{3/2}$  & 83$s_{1/2}$  & 72$p_{3/2}$  & 83$p_{1/2}$  & 5.291  & -7.29     & 12.04 \\
72$d_{3/2}$  & 67$s_{1/2}$  & 74$p_{3/2}$  & 66$p_{3/2}$  & 1.274  & 2.48      & 14.46 \\
72$d_{3/2}$  & 83$s_{1/2}$  & 73$p_{1/2}$  & 83$p_{1/2}$  & 2.753  & -23.51    & 21.44 \\
72$d_{5/2}$  & 84$s_{1/2}$  & 73$p_{3/2}$  & 84$p_{1/2}$  & 8.461  & -23.04   & 15.37 \\
74$d_{5/2}$  & 69$s_{1/2}$  & 76$p_{3/2}$  & 68$p_{1/2}$  & 2.276  & -8.37    & 16.99 \\
76$d_{3/2}$  & 68$s_{1/2}$  & 74$f_{5/2}$  & 68$p_{3/2}$  & 1.871  & -8.19    & 20.61 \\
76$d_{3/2}$  & 90$s_{1/2}$  & 75$f_{5/2}$  & 89$p_{3/2}$  & 5.491  & -21.47 &  19.85 \\
77$d_{3/2}$  & 90$s_{1/2}$  & 78$p_{3/2}$  & 90$p_{1/2}$  & 4.093  & -10.15    & 14.64 \\
78$d_{3/2}$  & 90$s_{1/2}$  & 79$p_{1/2}$  & 90$p_{3/2}$  & 11.745 & 32.52    & 15.76 \\
78$d_{5/2}$  & 91$s_{1/2}$  & 79$p_{3/2}$  & 91$p_{1/2}$  & 4.49   & -31.93    & 21.17 \\
79$d_{3/2}$  & 91$s_{1/2}$  & 80$p_{1/2}$  & 91$p_{1/2}$  & 5.394  & -34.26    & 19.43 \\
80$d_{5/2}$  & 95$s_{1/2}$  & 79$f_{7/2}$  & 94$p_{1/2}$  & 2.92   & 27.48     & 24.86 \\
81$d_{3/2}$  & 96$s_{1/2}$  & 80$f_{5/2}$  & 95$p_{1/2}$  & 0.434  & 27.78    & 47.1  \\
82$d_{3/2}$  & 97$s_{1/2}$  & 81$f_{5/2}$  & 96$p_{3/2}$  & 1.739  & -29.21   & 32.27 \\
83$d_{5/2}$  & 74$s_{1/2}$  & 81$f_{7/2}$  & 74$p_{1/2}$  & 1.601  & 12.11    & 23.11 \\
83$d_{3/2}$  & 77$s_{1/2}$  & 85$p_{3/2}$  & 76$p_{3/2}$  & 1.208  & 4.43     & 17.85 \\
84$d_{3/2}$  & 97$s_{1/2}$  & 85$p_{1/2}$  & 97$p_{3/2}$  & 1.023  & 44.04   & 39.34 \\
84$d_{5/2}$  & 98$s_{1/2}$  & 85$p_{3/2}$  & 98$p_{1/2}$  & 2.107  & -43.17  & 30.12 \\
85$d_{3/2}$  & 76$s_{1/2}$  & 83$f_{5/2}$  & 76$p_{3/2}$  & 1.239  & -12.97   & 27.56 \\
87$d_{5/2}$  & 103$s_{1/2}$ & 86$f_{5/2}$  & 102$p_{3/2}$ & 0.027  & 9.97    & 85.14 \\
88$d_{3/2}$  & 104$s_{1/2}$ & 87$f_{5/2}$  & 103$p_{3/2}$ & 0.428  & -38.88  & 56.63 \\
89$d_{3/2}$  & 104$s_{1/2}$ & 90$p_{3/2}$  & 104$p_{1/2}$ & 2.546  & -18.28    & 20.86 \\
90$d_{3/2}$  & 104$s_{1/2}$ & 91$p_{1/2}$  & 104$p_{3/2}$ & 8.088  & 58.38    & 21.69 \\
90$d_{5/2}$  & 105$s_{1/2}$ & 91$p_{3/2}$  & 105$p_{1/2}$ & 0.664  & -57.15   & 48.59 \\
91$d_{3/2}$  & 105$s_{1/2}$ & 92$p_{1/2}$  & 105$p_{3/2}$ & 4.749  & 60.88    & 26.27 \\
91$d_{5/2}$  & 106$s_{1/2}$ & 92$p_{3/2}$  & 106$p_{1/2}$ & 13.531 & -59.58    & 18.04 \\
91$d_{5/2}$  & 108$s_{1/2}$ & 90$f_{7/2}$  & 107$p_{1/2}$ & 9.242  & 46.39    & 20.16 \\
92$d_{3/2}$  & 106$s_{1/2}$ & 93$p_{1/2}$  & 106$p_{1/2}$ & 1.107  & -63.68 & 40.5  \\
93$d_{3/2}$  & 107$s_{1/2}$ & 94$p_{1/2}$  & 107$p_{1/2}$ & 10.237 & -66.34   & 19.56 \\
93$d_{3/2}$  & 109$s_{1/2}$ & 94$p_{3/2}$  & 109$p_{3/2}$ & 5.092  & 21.91     & 18.82 \\
93$d_{5/2}$  & 110$s_{1/2}$ & 92$f_{5/2}$  & 109$p_{3/2}$ & 2.305  & 13.05     & 21.23 \\
94$d_{3/2}$  & 84$s_{1/2}$  & 92$f_{5/2}$  & 84$p_{3/2}$  & 0.876  & -19.59  & 35.5  \\
94$d_{3/2}$  & 87$s_{1/2}$  & 96$p_{3/2}$  & 86$p_{3/2}$  & 0.986  & 7.36      & 22.61 \\
95$d_{5/2}$  & 88$s_{1/2}$  & 97$p_{3/2}$  & 87$p_{3/2}$  & 5.76   & 23.17     & 18.73 \\
97$d_{3/2}$  & 68$s_{1/2}$  & 100$p_{1/2}$ & 67$p_{3/2}$  & 4.506  & -3.42     & 10.23 \\
100$d_{3/2}$ & 93$s_{1/2}$  & 102$p_{1/2}$ & 92$p_{3/2}$  & 1.761  & 31.11    & 29.23 \\
101$d_{3/2}$ & 90$s_{1/2}$  & 99$f_{5/2}$  & 90$p_{1/2}$  & 4.28   & 26.23    & 21.55 \\
102$d_{3/2}$ & 95$s_{1/2}$  & 104$p_{1/2}$ & 94$p_{1/2}$  & 2.67   & -33.7    & 24.43 \\
103$d_{3/2}$ & 68$s_{1/2}$  & 100$f_{5/2}$ & 68$p_{3/2}$  & 0.232  & 3.2      & 30.21 \\
103$d_{3/2}$ & 68$s_{1/2}$  & 100$f_{5/2}$ & 68$p_{3/2}$  & 0.232  & 3.2      & 30.21 \\
103$d_{3/2}$ & 92$s_{1/2}$  & 101$f_{5/2}$ & 92$p_{3/2}$  & 0.65   & -28.46   & 44.41 \\
105$d_{3/2}$ & 97$s_{1/2}$  & 107$p_{3/2}$ & 96$p_{3/2}$  & 0.776  & 11.53    & 28.45 \\
106$d_{5/2}$ & 98$s_{1/2}$  & 108$p_{3/2}$ & 97$p_{3/2}$  & 2.873  & 36.14    & 27.39 \\
107$d_{3/2}$ & 99$s_{1/2}$  & 109$p_{3/2}$ & 98$p_{1/2}$  & 0.305  & -12.44   & 37.25 \\
109$d_{5/2}$ & 97$s_{1/2}$  & 107$f_{7/2}$ & 97$p_{1/2}$  & 5.169  & 36.87    & 22.67 \\
110$d_{3/2}$ & 98$s_{1/2}$  & 108$f_{5/2}$ & 98$p_{1/2}$  & 1.111  & 37.18    & 37.95
\end{tabular}
\end{ruledtabular}
\end{table*}

\begin{table*}
\caption{ \label{table_KK}A selection of K-K pair states with F\"orster defects $|\Delta|/(2\pi) < 12\,$MHz and $|C_{3k}|> 1\,$GHz$\micm^3$ for principal quantum numbers $50 \leqslant n \leqslant 100$. Values for Rb-Rb, Rb-Cs and Cs-Cs can be found in Table~II, III and IV of Ref.~\cite{Beterov2015}. We find an interaction strength of \SI{2.3}{\MHz} at ${R = \SI{19}{\micm}}$ for $n <$ 90 which is very similar to the \SI{2}{\MHz} at ${R = \SI{20}{\micm}}$ reported for Rb-Cs. }
\begin{ruledtabular}
\begin{tabular}{cccccccc}
 $\text{K initial} $& $\text{K final}$ & $\text{K initial}$ & $\text{K final}$ & \text{$\Delta/2 \pi $ (MHz)}&  \text{$C_{3k}$ (\SI{}{\GHz\: \um^3})} & \text{$R_c$ (\SI{}{\um})}\\
$56s_{1/2}$ &$56p_{1/2}$ &$59s_{1/2}$ &$58p_{1/2}$ & 10.21 & -3.70  & 8.8  \\ 
$59s_{1/2}$ &$59p_{3/2}$ &$62s_{1/2}$ &$61p_{1/2}$ & -0.27 & 4.55&  32.1 \\ 
$62s_{1/2}$ &$62p_{3/2}$ &$65s_{1/2}$ &$64p_{3/2}$ & -7.22 & -5.59  & 12.9 \\ 
$64s_{1/2}$ &$64p_{3/2}$ &$95s_{1/2}$ &$93p_{3/2}$ & 10.80 & 1.90 & 7.9 \\ 
$66s_{1/2}$ &$66p_{3/2}$ &$98s_{1/2}$ &$96p_{1/2}$ & -1.69 & -2.17 & 13.7 \\ 
$78s_{1/2}$ &$78p_{1/2}$ &$82s_{1/2}$ &$81p_{1/2}$ & -0.47 & -14.49  & 38.6 \\ 
$81s_{1/2}$ &$81p_{1/2}$ &$85s_{1/2}$ &$84p_{3/2}$ & 4.34 & 16.94  & 19.8 \\ 
$82s_{1/2}$ &$82p_{3/2}$ &$86s_{1/2}$ &$85p_{1/2}$ & -1.29 & 17.65  & 30.1 \\ 
$83s_{1/2}$ &$83p_{3/2}$ &$87s_{1/2}$ &$86p_{1/2}$ & -10.23 & 18.53  & 15.4 \\ 
$85s_{1/2}$ &$85p_{3/2}$ &$89s_{1/2}$ &$88p_{3/2}$ & 5.50& -20.48 &  21.7 \\ 
$86s_{1/2}$ &$86p_{3/2}$ &$90s_{1/2}$ &$89p_{3/2}$ & -2.19 & -21.46  & 30.0 \\ 
$87s_{1/2}$ &$87p_{3/2}$ &$91s_{1/2}$ &$90p_{3/2}$ & -9.19 & -22.48  & 18.9 \\ 
 \\
 \end{tabular}
\end{ruledtabular}
\end{table*}
\clearpage
\twocolumngrid

%


\end{document}